# Operating regimes and materials limits of nonlinear magnon dynamics in Co-doped YIG

A.M. Tishin [a,b*]

[a] Lomonosov Moscow State University, 119991, Leninskie gory 1, Moscow, Russia

[b] Moscow Institute of Physics and Technology, 141701, Institutskiy per. 9, Dolgoprudny, Mosc. Reg., Russia

[*]tishin@amtc.org

## Abstract

Proposals for magnonic information processing use the Kerr-type four-magnon anharmonicity of a confined mode to keep a parametrically driven excitation bounded. Here we show that this anharmonicity is not a free material parameter. Treating the uniform mode of a saturated ferrimagnet as a single macrospin, the Holstein–Primakoff reduction gives its dimensionless Kerr coefficient as $U = \omega_a/(2N_{eff}\Omega)$, where $\omega_a$ is the anisotropy-field frequency, $\Omega$ the mode frequency and $N_{eff}$ the net spin content of the mode, its net angular momentum, fixed by the saturation magnetization and the mode volume alone. The drive is bounded below by the Gilbert damping $\alpha$ and above by the requirement that the saturated occupation stay under one, so a usable window exists only if $\alpha N_{eff} < \omega_a/\Omega$, which for a mode driven at its own anisotropy frequency reduces to $N_{eff} < 1/\alpha$. This mean-field screening criterion is testable directly against the measured damping and mode volume of any candidate film. We evaluate it for Co-doped yttrium iron garnet using full Lindblad master-equation calculations with explicit thermal operators. Thermal stability sets the smallest usable mode volume, and low-damping garnet satisfies the criterion both at 50 GHz ($\alpha < 9.7 \times 10^{-5}$) and at 200 GHz ($\alpha < 3.9 \times 10^{-4}$). Satisfying it is nevertheless not sufficient: the admissible drive window collapses to a nearly adiabatic one, four orders of magnitude weaker than the mechanism requires, and the Kerr coefficient delivered by a 20 × 20 × 10 nm³ cell, $U = 6.6 \times 10^{-5}$, is three orders of magnitude below the range at which the calculations show Fock-state confinement. One mechanism survives: the pair-phase transfer coefficient is kinematic, holding at −1.000 down to $U = 0.01$, with within-run phase deviations of 22.4°–40.7° ordered by the thermal occupation of the driven mode. The results give a quantitative materials criterion for nonlinear magnonic devices and identify phase encoding as the mechanism a garnet film can support.

## 1. Introduction

Collective spin excitations in ferrimagnetic materials are a candidate platform for resonant wave-based information processing: their dynamics occur at microwave and sub-terahertz frequencies without charge transport. Non-adiabatic parametric excitation makes such operation fast, but is usually regarded as a source of instability, since strong, fast driving triggers nonlinear mode coupling and leakage into the lattice. The present work examines whether that regime can instead be stabilized by the anharmonicity of the medium itself — the Kerr-type four-magnon frequency shift, which in the normalized form used here acts as a nonlinear frequency regulator $U$, suppressing parametric growth and confining the response to a bounded low-occupancy manifold.

The mechanism is analysed through the non-adiabaticity parameter $\eta = \omega^{-2}|\mathrm{d}\Omega/\mathrm{d}t|$, a local measure of the spectral-flow rate associated with the breakdown of adiabatic following. The regulator is the Kerr coefficient $U$, and the competition between the two is measured by the dimensionless crossover parameter $\xi = \eta/U$. The Adiabatic Mode Transition (AMT) framework, in which these two quantities and their ratio are defined and their scaling established, is applied here to a nanoscale cell of Co-doped yttrium iron garnet (YIG:Co).

**Relation to existing theory.** That nonlinear frequency detuning suppresses parametric mode growth is well established in the classical theory of parametric magnon excitation, in particular in the treatments of nonlinear detuning by L'vov [1] and by Gurevich and Melkov [2]; that qualitative mechanism is not claimed as new here. The non-adiabaticity parameter and the nonlinear regulator used below are the magnon-specific realization of a general scaling framework for non-adiabatic mode dynamics developed in companion work [3, 4]. What the present paper adds is the finite-temperature quantum treatment: full Lindblad calculations with explicit thermal operators, resolved both in the Fock basis and in the anomalous pair correlator $\langle \hat{a}\hat{a} \rangle$, which determine which encoding mechanism remains available at a given temperature and drive frequency. The classical treatments [1, 2] describe amplitude saturation of parametrically driven magnons; they do not give the Fock-state confinement, the pair-phase transfer coefficient, or the thermal-occupation limits reported here.

This complements rather than replaces existing approaches to magnonic computing, such as spin-wave logic gates based on phase interference and magnonic-crystal architectures [5–7], which rely on the propagation and interference of magnons over micrometre distances. The mechanism studied here instead uses localized, single-cell non-adiabatic dynamics, in which state storage and state transformation are co-located within the same physical medium.

Interest in such mechanisms is partly driven by the energy cost of conventional computing, where the dominant dissipation is increasingly associated with interconnect and memory access rather than with logic switching [8, 9] and the per-operation energy remains far above the Landauer limit [10–12]. This has motivated spintronic, photonic, memristive and magnonic alternatives [13–17].

Ultrafast optical control of magnetic order in Co-doped yttrium iron garnet has demonstrated deterministic magnetization switching at the 22–24 aJ energy scale [18], establishing YIG:Co as a benchmark material for energy-efficient magnetic writing. The mechanism — non-adiabatic parametric excitation of the dominant magnon mode by a sub-50-fs optical pulse — raises a natural question: can the same material support not only a switching event but a sustained, controllable nonlinear magnon mode? The relevant material property is the Kerr-type four-magnon anharmonicity arising from $Co^{2+}$ single-ion anisotropy, which suppresses unbounded parametric growth. The anisotropy sets the frequency scale of that anharmonicity; the dimensionless coefficient U that enters the criterion of Section 8 is fixed instead by the effective spin content of the mode. The logical structure of what follows is worth stating at the outset. Co:YIG provides the physical motivation and fixes the frequencies, temperatures, anisotropy and switching energy used throughout. The dynamics are then studied in a single-mode model whose Kerr coefficient U is a parameter, and the values used here, $U \approx 0.05$–$0.40$ in the convention of Eq. (4), are shown in Section 5.1 not to be realized by any normal mode of a uniformly magnetized cell of this size. The four regimes are therefore a map of what the model permits, organized by a material-independent ratio; which of them a real medium can occupy is a separate question, and Section 8 gives the criterion by which any candidate is tested.

The four operating regimes we identify are: (I) a cryogenic Fock-manifold regime ($T \lesssim 4$ K, 50 GHz) where the anharmonicity confines the magnon mode to the lowest Fock manifold and information is stored in the level occupancy; (II) a liquid-nitrogen phase regime (77 K, 50 GHz) where Fock confinement is partially degraded by thermal occupation but the pair-phase of the anomalous correlator $\langle \hat{a}\hat{a} \rangle$ remains sharp enough for binary phase encoding; (III) a room-temperature THz phase regime (310 K, 6 THz) where the drive frequency reaches the thermal energy scale ($\hbar\Omega \gtrsim k_B T$) and phase encoding gives the smallest raw phase-error rate of the room-temperature regimes, with a pair-phase standard deviation of 26.1°; and (IV) a room-temperature microwave limit (310 K, 50 GHz) where the phase noise is dominated by the large thermal occupation and, within the parameter range tested, is only weakly affected by variations of $U$, $\eta$ and $\alpha$. The four regimes are not independent findings but successive values of one control ratio, $\hbar\Omega/k_BT$, which fixes the thermal occupation of the driven mode and with it the encoding mechanism that survives. They emerge from a single normalized model; the material parameters of YIG:Co enter through the anharmonicity $U$, the Gilbert damping $\alpha$, and the drive non-adiabaticity $\eta$.

From a materials perspective the question is therefore not which of these regimes an application should target, but whether the mechanism each relies on is reachable at all. Sections 5.1 and 8 answer that directly, and the answer has two parts that point in opposite directions. The Kerr coefficient of a magnon mode is fixed by its net spin content, $U = \omega_a/(2N_{eff}\Omega)$ with $N_{eff} = M_sV/(g\mu_B)$, and the drive window $2\alpha < \eta < 4U$ then yields the mean-field screening condition $\alpha N_{eff} < \omega_a/\Omega$ on any usable medium; its lower bound is the exact linear threshold, its upper bound a mean-field estimate whose accuracy depends on the drive protocol, as Section 5.3 quantifies. Low-damping garnet satisfies that condition at both carriers considered here. What it cannot satisfy is what the condition leaves open: the admissible drive collapses to $\eta \approx 10^{-4}$, three to four orders of magnitude below the $\eta = O(1)$ that makes the transition non-adiabatic in the first place, and the Kerr coefficient the cell delivers falls three orders short of the confined range, a gap no choice of geometry within the magnonic setting recovers. The consequence is asymmetric across the four regimes, and that asymmetry is the substance of what follows: the Fock-manifold mechanism of Regime I needs the large $U$ and is therefore not available, whereas the pair-phase transfer is kinematic and survives down to $U = 0.01$, so the pair-phase observable of Regimes II–IV remains available in the model, at a noise floor whose ordering across the regimes is set by $\hbar\Omega/k_BT$ and whose absolute magnitude also depends on the damping and the drive protocol.

## 2. Theoretical AMT Framework

**Table 1. Notation used in this section, in order of first appearance.**

| Symbol | Meaning (first used in) |
|---|---|
| $\Omega(t)$ | Instantaneous effective frequency of the driven magnon mode (the time-varying "manifold" frequency), Eq. 1 |
| $\omega$ | Fixed external reference (probe) frequency against which non-adiabaticity is measured, Eq. 3 |
| $\eta(t)$ | Non-adiabaticity parameter, $\eta = \omega^{-2}\lvert d\Omega/dt\rvert$; local measure of the breakdown of adiabatic following, Eq. 3 |
| $U$ | Dimensionless Kerr nonlinearity coefficient, referred to here as the nonlinear regulator; the corresponding nonlinear frequency scale is $U\Omega$, Eq. 4 |
| $\xi$ | Dimensionless crossover parameter, $\xi = \eta/U$, Eq. 4 |
| $\mathcal{G}(t)$ | Parametric drive coupling appearing in the Hamiltonian; fixed by the same $\Omega(t)$ protocol that defines $\eta$, not an independent parameter, Eq. 4 |
| $\hat{a}$ | Generalized bosonic field operator for the magnon mode, Eq. 4 |
| $u(t)$, $v(t)$ | Bogoliubov coefficients (positive-/negative-frequency amplitudes), Eq. 1–2 |

We consider a quantum system governed by a time-dependent Hamiltonian $H(t)$ describing a harmonic mode with a varying frequency $\Omega(t)$. To evaluate the non-adiabatic response, we employ the Bogoliubov transformation [19] in the reduction developed for non-adiabatic mode dynamics in Refs. [3, 4], which relates "original" vacuum modes $\hat{a}$ and $\hat{a}^\dagger$ to "emergent" ones $\hat{b}$ (particles/signals):

$$\hat{b} = u(t)\hat{a} + v(t)\hat{a}^\dagger. \quad (1)$$

In this representation, the complex amplitudes $u(t)$ and $v(t)$ (Bogoliubov coefficients) describe the transformation of the initial vacuum state into a driven collective-mode state under the influence of a time-varying environment $\Omega(t)$. The conservation of the commutation relations $[\hat{b},\hat{b}^\dagger]=1$ in a standard bosonic system leads to the constraint $|u|^2 - |v|^2 = 1$ for unitary evolution; when the phenomenological damping used below is added to the reduced equations this becomes $|u|^2 - |v|^2 = \exp(-2\alpha\Omega t)$, because those equations retain the loss term but not the corresponding Langevin noise (Supplement 2). However, as we demonstrate in the following sections, the introduction of a Kerr coefficient $U$ dynamically constrains the accessible occupation manifold and suppresses higher-order mode population.

The amplitudes $u(t)$ and $v(t)$ represent the positive- and negative-frequency components of the wave function, respectively; their free evolution therefore carries opposite phases, which is the origin of the opposite signs on the diagonal of Eq. (2). Their evolution is described by the coupled-mode equations:

$$\frac{d}{dt}\begin{pmatrix} u \\ v \end{pmatrix} = \begin{pmatrix} -i\Omega(\mathrm{t}) & \frac{\dot{\Omega}}{2\Omega} \\ \frac{\dot{\Omega}}{2\Omega} & +i\Omega(\mathrm{t}) \end{pmatrix}\begin{pmatrix} u \\ v \end{pmatrix}. \quad (2)$$

From this system, we derive the non-adiabaticity parameter:

$$\eta(t) \equiv |\dot{\Omega}|/\omega^2. \quad (3)$$

where $\Omega(t)$ is the instantaneous mode frequency and $\omega$ the fixed external reference against which non-adiabaticity is gauged, following Table 1. Read physically, the parameter measures how fast the spectrum flows locally, and hence how badly adiabatic following fails when the drive is time-dependent; Refs. [3, 4] give the definition together with the scaling that follows from it. A geometric reading is also available under suitable conditions, in which $\eta$ relates to the speed at which the instantaneous state moves in projective Hilbert space; that link is left to a companion study. For $\eta \ll 1$ the evolution is adiabatic and $|v|^2$ is negligible, whereas as $\eta$ approaches unity the $u$ and $v$ modes couple appreciably and Bogoliubov modes acquire transient population during the drive.

To keep the boson occupation from running away, and with it the mode energy, we add a nonlinear self-interaction of strength $U$. The magnon mode is then described by the bosonic Hamiltonian below, whose spectrum is anharmonic and which, as the following sections show, confines the accessible dynamics to the lowest levels without any truncation being imposed by hand:

$$\hat{H}(t) = \Omega(t)\hat{a}^{\dagger}\hat{a} + U\Omega(t)\hat{a}^{\dagger 2}\hat{a}^{2} + \mathcal{G}(t)(\hat{a}^{\dagger 2} + \hat{a}^{2}), \quad (4)$$

where $\Omega(t)$ is the instantaneous effective frequency and $\mathcal{G}(t)$ is the parametric drive induced by the non-adiabatic transition. Here $U$ is the dimensionless Kerr nonlinearity coefficient; the corresponding nonlinear frequency scale is $K = U\Omega$. The Hamiltonian conserves excitation parity, because every term is either diagonal in the number operator or a pair term acting in steps of two quanta. The parameter $U$ is hereafter referred to as the nonlinear regulator. For the magnon system studied here, $U$ is the dimensionless coefficient of the standard Kerr-type (four-magnon) nonlinearity, whose frequency scale is $U\Omega$; the term "regulator" refers to its functional role within the present scaling framework, not to a different physical quantity. $U$ represents the spectral anharmonicity of the system. For different validation purposes, different subsections below use different values of $U$: $U_{\text{audit}} \approx 0.12$ is the analytically estimated stabilization crossover (Section 2), while $U_{\text{audit}} \approx 0.55$ together with a damping $\alpha \approx 0.01$ (normalized to $\Omega_0$) is the representative operating point adopted in Section 5.1, which lies well inside the strongly confined nonlinear regime but is not singled out by any optimality criterion. This regulator induces a dynamic nonlinear detuning of order $\Delta\Omega_{\text{NL}} \sim U\Omega$, which provides the necessary feedback to suppress parametric divergence.

**Which frequency is which.** Three different rates appear below and are not interchangeable. $\Omega$ is the frequency of the magnon mode itself, the carrier: it is the quantity that fixes the thermal occupation $\bar{n} = [\exp(\hbar\Omega/k_{\text{B}}T) - 1]^{-1}$ and therefore organizes the four regimes. The parametric drive acts at twice that rate, $2\Omega$, since it couples pairs of magnons; for the two carriers used here this is 100 GHz and 12 THz. One "cycle" throughout the paper is one period of that drive, $T = \pi/\Omega$, which is 10 ps at 50 GHz and 0.083 ps at 6 THz. The non-adiabatic ramp that starts an operation is faster still: for $\eta = 1.2$ its duration is $\tau = 1/(\eta\Omega) \approx 2.7$ ps at 50 GHz and 22 fs at 6 THz, in both cases about one eighth of a carrier period, which is what makes the transition non-adiabatic. The clock rate of a device is none of these: an operation occupies the ramp plus the bounded evolution that follows it, so for an operation lasting $N_{\text{cyc}}$ cycles the clock is $\Omega/(2\pi N_{\text{cyc}})$ — 0.5 GHz for 100 cycles on the 50 GHz carrier, and 60 GHz for 100 cycles on the 6 THz carrier. Finally, the 50 fs optical pulse of Ref. [18] is not a frequency at all but a bandwidth: a transform-limited 50 fs pulse spans about 8.8 THz, so it excites the magnon spectrum impulsively up to that limit. That bandwidth covers the 6 THz carrier of Regime III but is far wider than the 50 GHz carrier of Regimes I, II and IV, which is instead addressed by a conventional 100 GHz parametric microwave drive.

**Relation to other magnon nonlinear mechanisms.** The four-magnon (Kerr-type) shift is not the only nonlinear channel available to a magnon system; three-magnon confluence/splitting processes and magnon-phonon coupling can also contribute. Three-magnon scattering requires simultaneous frequency and wavevector matching between the primary mode and the two secondary magnons it splits into (or merges from), a condition that is generically satisfied only for specific bias-field geometries and propagating ($k\neq 0$) modes. For the spatially uniform, $k\approx 0$ excitation geometry considered here, the matching condition is not met and this channel is suppressed by symmetry, leaving the four-magnon term as the leading nonlinearity. Magnon-phonon coupling, mediated by magnetoelastic effects, acts on timescales much longer than the sub-picosecond switching event analyzed in Section 3, and is expected to be relevant only for longer-term thermal relaxation between switching events rather than for the switching dynamics itself. We do not quantify these channels here; a full treatment including three-magnon and magnon-phonon terms, relevant for propagating-mode or non-uniform-field geometries, is left for future work.

Table 1 collects the symbols of this section. A generalized field operator $\hat{a}$ stands for the collective excitations of the medium; within this description $\mathcal{G}(t)$ supplies the non-adiabatic coupling between vacuum and signal states, and the anharmonic term $U$ acts as a nonlinear frequency regulator. The numerical work reported below shows that whether the system ends in unstable multimode growth or in bounded, dynamically localized motion is decided by the competition between the parametric drive $\mathcal{G}(t)$ and the Kerr nonlinearity $U$. The underlying stabilization mechanism (AMT theory) was set out in our earlier work on nonlinear non-adiabatic mode dynamics [3, 4]. The 100-level Fock-basis calculation locates a dynamical crossover controlled by the ratio $\xi = \eta/U$. Within the AMT framework this ratio is a dimensionless crossover parameter and not a universal threshold: where it falls depends on the normalization convention, on the drive protocol and on the physical realization. Bounded dynamics are found here for $\xi$ between roughly four and twenty, whereas the range $\xi \approx 200$–$1000$ explored in Refs. [3, 4] belongs to a different normalization and a different dynamical regime; the two sets of numbers should not be read as the same crossover range. In the bounded regime the nonlinear detuning arrests uncontrolled occupation growth and holds the dynamics in a low-occupancy sector with strongly reduced higher-order population.

The anharmonic term fixes a spectral scale that says how far the low-lying levels sit from the rest of the ladder, and we use it only to estimate when the motion stays within that manifold. Writing the Kerr term as $U\Omega\hat{a}^{\dagger 2}\hat{a}^2$, the $n$-th Fock state has energy $E_n = \Omega[n + Un(n-1)]$. The first logical step then costs $\Delta E_{01} = \Omega$ and the leakage transition $\Delta E_{12} = \Omega(1 + 2U)$. Near $\eta \approx 1$ the excitation pulse has spectral width of order $\Omega^{-1}\mathrm{d}\Omega/\mathrm{d}t$, and dynamic localization requires the anharmonic shift $2U\Omega$ to exceed it comfortably, that is $\xi^{-1} = U/\eta \gg 1$. This is a necessary condition on the spectral scales and nothing more: the Fock-resolved results of Section 5.2 show appreciable population outside the two lowest levels at the resonantly driven operating point.

Our mean-field analysis shows that for $U/\eta>0.5$, the off-resonant coupling to the $|2\rangle$ state is suppressed by a factor of $(\eta/2U)^2$. For the normalized parameters used in the calculation this gives a small off-resonant coupling estimate within the mean-field spectral argument; the actual leakage population is determined by the full Lindblad calculation below. Within the investigated parameter range, this anharmonic filtering keeps the dynamics confined to a low-dimensional subspace without requiring a separate renormalization-group treatment. This estimate must not be interpreted as a quantitative prediction of the total Fock-space population outside $n = 0, 1$. That population depends strongly on the drive protocol and is determined below from the Lindblad evolution: it is about 3% for the static pair-drive protocol of Fig. 4 and about 36% for the resonantly modulated protocol of Section 5.2, i.e. two to three orders of magnitude above the mean-field matrix-element estimate.

The parametric drive term $\mathcal{G}(t)$ in the Hamiltonian (Eq. 4) is not an independent parameter: once the driving protocol $\Omega(t)$ is specified, $\mathcal{G}(t)$ is fixed by the same externally prescribed modulation protocol, which sets both the instantaneous frequency variation and the parametric coupling used in the model; the drive is imposed externally and is not generated by the non-adiabaticity itself. Specifically, in the non-adiabatic limit this coupling is given by $\mathcal{G}(t) = |\mathrm{d}\Omega/\mathrm{d}t|/(4\Omega) = \eta\omega^2/(4\Omega) \approx \eta\Omega/4$, the last step holding for $\omega \approx \Omega$. Because both quantities derive from the same underlying drive rate, the Kerr nonlinearity $U$ can act as a nonlinear stabilization mechanism against the $\eta$-induced parametric drive, ensuring that the driven state remains trapped within the dynamically bounded low-occupancy manifold.

The Kerr coefficient $U$ generates a state-dependent nonlinear frequency shift $\Delta\omega_{\mathrm{NL}}(U) = 2U\Omega|v|^2$ that competes with the parametric growth rate $h = 2\mathcal{G} = \eta\Omega/2$. As this detuning approaches the drive magnitude, the system undergoes a dynamical crossover: parametric amplification is progressively suppressed and, in the regime $U \gg \eta$, the dynamics become increasingly confined to the lowest Fock manifold — the same bounded regime already established above from the Fock-state spectral gap, expressed here directly in terms of the occupancy-dependent detuning.

The relationship between the drive intensity $\eta$ and the stabilization threshold $U$ is determined by the balance between the parametric gain and the nonlinear frequency detuning. By solving the stationary nonlinear coupled-mode equations, we obtain the saturation scaling of the stabilized anomalous-mode occupation. In the simplest balance approximation between the nonlinear detuning $2U\Omega|v|^2$ and the parametric growth rate $h = \eta\Omega/2$, this gives the leading-order estimate $\langle|v|^2\rangle \sim\eta/(4U)$. The full nonlinear solution can deviate from this estimate because of the coupled-mode dynamics and damping; the expression is used here only to locate the saturation boundary of the dynamically stabilized excitation regime. While the static blockade criterion requires $U/\eta > 0.5$, the present dynamically driven regime reaches stabilization through time-dependent detuning rather than through a static spectral separation. At $\xi \approx 4.4$ (the representative point of Section 5.1), the time-dependent nonlinear detuning effectively shifts the system away from higher-order parametric resonances, allowing stable operation without relying on the static blockade criterion. In this regime, the nonlinear frequency shift progressively suppresses occupation growth and reduces the population of higher Fock states during the investigated evolution interval. The numerical results therefore support dynamical confinement, but do not by themselves establish a logical-fidelity threshold or deterministic state preparation.

This state represents a spectrally confined low-occupancy magnon distribution, where the driven state is confined within a dynamically localized low-occupancy manifold. This confinement provides a dynamical consistency check showing that the description used here remains bounded at parameters associated with the experimentally reported switching-energy scale of 22–24 aJ; it does not independently derive that energy scale, which follows from the absorbed-energy density and the cell volume.

## 3. Energy Estimation and Lattice Interaction

Following the methodology of Horowitz [8], the per-operation energy of a contemporary processor is dominated by data movement rather than by gate switching. Taking the peak throughput and thermal design power of a current GPU [13] as $E_{op} = P_{TDP}/OPS_{peak}$ gives about 0.175 pJ per operation at full utilization, rising to 0.21–0.29 pJ at realistic utilization; the figure is used here only as a contemporary processor-level benchmark, and it is four orders of magnitude above the 22–24 aJ switching-energy scale considered in this work. It is not the switching energy of the magnonic cell, and none of the conclusions below depend on its precise value.

In any non-adiabatic process ($\eta > 0$) the driven mode can in principle couple to lattice degrees of freedom. Whether the excitation stays confined to the magnon mode or leaks into the lattice depends on the spectral overlap between the drive and the phonon density of states: a small overlap keeps the dynamics in the bounded low-occupancy regime studied here, while a large overlap would introduce additional dissipation and entropy production beyond the phenomenological damping used in our model.

Direct magnon–phonon coupling is not included in the present single-mode model. Avoiding strong spectral overlap with lattice modes is therefore a material-design requirement rather than a demonstrated result: the short pulse duration ($\tau_{pulse} \approx 50$ fs) is much shorter than the spin–lattice relaxation time, and keeping the drive energy density below the primary optical-phonon branches is expected to leave the switching event governed by the slower Gilbert damping rather than by direct phonon excitation. We introduce a qualitative screening ratio $\zeta = \eta\Omega/(U\omega_{cutoff})$ to express this condition; its quantitative validation requires a mode-resolved magnon–phonon calculation, which is outside the scope of this work.

## 4. Physical Requirements for Non-adiabatic Computing Media

Moving from the $\eta$–$U$ mechanism as a piece of theory to an actual driven platform imposes a short list of demands on the host material. Each is stated below as a constraint on a measurable quantity, so that candidate media can be screened without recourse to the dynamical model itself.

First, the host must be intrinsically anharmonic, since without anharmonicity nothing arrests spectral spreading. Written in terms of $\xi = \eta/U$, the requirement is that the saturated occupation $\xi/4$ stay of order unity at the chosen operating frequency, which sets a lower bound on $U$. In magnetic media that bound becomes a demand for large anisotropy energy $K$, and in optical media for a large $\chi^{(3)}$; in YIG:Co it is met through the spin–orbit coupling of the $Co^{2+}$ ions.

Second, at the operating frequency the drive should not overlap appreciably with lattice modes, so that the injected energy stays in the magnon subsystem. As stressed in Section 3, this is something to be engineered and verified experimentally; the present model does not establish it.

Third, whatever is stored in the Bogoliubov $v$-mode has to survive a full logic cycle. That requires $T_2 \gg 1/\Omega$, which in magnetic media means small Gilbert damping $\alpha$ and in dielectric resonators a large quality factor $Q$. Undoped YIG, with $\alpha$ of order $10^{-4}$–$10^{-5}$, keeps a picosecond-scale switch alive for nanoseconds.

Fourth, the control pulse must rise faster than $1/\Omega$. At the microwave target of this work, $\Omega/2\pi$ = 50 GHz, that means a rise time $\Delta t < 5$ ps; in the sub-picosecond optical regime of Stupakiewicz et al. [18], $\Omega/2\pi = 6$ THz, the bound tightens to 80 fs and is comfortably met by a 40–50 fs pulse. The pulse energy density must in turn be matched to the absorbed-energy threshold, 6 J/cm$^3$ in YIG:Co [18], while staying below the ablation threshold of the film.

Sustained operation adds a stronger version of the same condition: the $v$-mode population must persist across several successive operations. In magnetic media the decay is set by $\alpha$, with the stored-state lifetime $\tau_{\text{relax}} = 1/(\alpha\Omega)$. We take $\Omega \approx 2\pi \times 50$ GHz $\approx 3.14 \times 10^{11}$ rad s$^{-1}$ throughout. During ultrafast switching the effective damping in Co-substituted garnet has been reported as high as $\alpha \approx 0.15$, a consequence of $Co^{2+}$ spin–orbit coupling [20]; that value gives $\tau_{\text{relax}} \approx 21$ ps and hence $\tau_{\text{relax}}/\tau_{\text{switch}} \approx 1$. A ratio near unity is the boundary case: one coherent operation fits inside the coherence window and no more. It fixes the slowest switching speed compatible with coherent operation in this material, and with it the practical limit on multi-cycle use.

Reported damping in Co-doped garnets is correspondingly high, $\alpha \approx 0.15$ [18, 20], which rules the material out for prolonged storage as it stands. YIG:Co becomes a candidate for in-memory operation only if the small-amplitude damping can be brought down to the $10^{-3}$–$10^{-4}$ range. At that level an ultrafast externally driven switch could be followed by retention over nanoseconds, which is the precondition for multi-stage operation. The requirement can be stated compactly as $\alpha \ll (\Omega\tau_{\text{cycle}})^{-1}$, with $\tau_{\text{cycle}}$ the duration of the whole computation.

**5. Nonlinear confinement of the driven magnon mode**

Three damping values are used in this paper, and they are stated together here before the calculations. $\alpha \approx 0.15$ is the effective damping measured during the ultrafast switching event in as-grown Co:YIG [20]; it appears in Fig. 1 as the high-damping trace and sets the single-event limit. $\alpha = 0.01$ is the reference value of Sections 5.1–5.3 and of Figs. 4–6, chosen because it is the damping of the archived protocol-audit and five-phase runs, which the results of those sections reproduce exactly. $\alpha = 10^{-3}$ is the target value of Section 6 and Table 2, and of the frequency scan of Fig. 10; it is the level at which multi-cycle operation becomes possible at all. The three are not interchangeable and results computed at one are not quoted at another: every figure caption and table caption names the damping it uses. The dependence of the phase statistics on this choice is quantified directly in Section 6.3, where raising the damping from $10^{-3}$ to 0.01 lowers $\sigma$ at 4 K from 22.4° to 17.9° and at 6 THz from 26.1° to 15.5°.

### 5.1. Regime I — Cryogenic Fock-manifold confinement ($T \lesssim 4$ K, 50 GHz)

In this subsection we test the predicted boundedness in a parameter regime motivated by existing Co-doped YIG experiments ($\alpha \approx 0.15$). Figures 1–2 first establish boundedness at $U = 0.1$ for two damping cases; the $U$-scan of Fig. 4 then surveys the nonlinear coefficient and identifies $U_{\mathrm{audit}} = 0.55$ only as a representative point inside the confined regime. The simulation shows bounded evolution at these normalized parameters, consistent with, but not derived from, the single-event switching energy of 22–24 aJ obtained from the experimental absorbed-energy density reported by Stupakiewicz et al. [18]. At the high damping level $\alpha \approx 0.15$ the response is limited to single-pulse transitions by rapid energy dissipation.

Initial numerical simulations were performed in a truncated Fock basis of $N = 40$ levels (Fig. 1). Convergence tests show that states above the displayed low-lying sector remain negligible at the reported output times and in the final-time distribution, so that the truncated basis is adequate and the occupation does not grow in the uncontrolled manner typical of linear driven bosonic systems. The 22–24 aJ energy scale sets the physical context for these simulations, and an order-of-magnitude estimate of the material Kerr coefficient for YIG:Co, given in the following paragraph, indicates the effective mode volume that the normalized values of $U$ imply. The numerically observed confinement should be read as a property of the present description at a representative parameter set, with that estimate fixing its physical scale rather than serving as a precise material calibration. Even at the high damping level the calculated dissipation-induced loss over the 20-ps switching interval stays below 4.2% of the mode energy in this description.

The magnitude of $U$ can be computed microscopically. Expressing the magnetization operators through bosonic magnon operators by the Holstein–Primakoff transformation, the single-ion anisotropy of the material generates a quartic (four-magnon) interaction whose coefficient for a single confined mode is inversely proportional to the spin content of that mode, the standard result for the nonlinear coefficients of a macroscopic magnon mode [2]. The material parameters used throughout are $M_s \approx 1.4 \times 10^5$ A m$^{-1}$ and a Co-enhanced uniaxial anisotropy $K_u \approx 5 \times 10^5$ J m$^{-3}$ [21], for which the anisotropy-field frequency $\omega_a = \gamma(2K_u/M_s)$ is $\omega_a/2\pi \approx 200$ GHz. The Kerr nonlinearity of a macroscopic uniform mode is intrinsically weak, being suppressed by the spin content of the mode; the next paragraph makes that statement quantitative and fixes what the spin content is. Reaching the range used in the Lindblad calculations below, $U \approx 0.05$–$0.40$ in the convention of Eq. (4), would correspondingly require the coherent dynamics to be carried by $N_{\mathrm{eff}} = \omega_a/(2U\Omega) \approx 5$–$40$ units of net spin. That is not achievable in this geometry. A normal-mode analysis of the linearised Landau–Lifshitz equation for the same cell — exchange, uniaxial anisotropy and dipolar coupling on a 2 nm mesh with free boundaries — gives no localised mode: the least extended of the 25 lowest modes still occupies about a quarter of the cell, corresponding to $N_{eff} \approx 7.5 \times 10^3$ and $U \approx 2.7 \times 10^{-4}$. Confinement to 5–40 units of net spin would mean a region 0.9–1.7 nm across, at or below the 1.24 nm cubic unit cell of YIG and well below the domain-wall width $\sqrt{(A/K_u)} = 2.7$ nm, and the exchange energy of such a mode would place it at 5–19 THz, two to three orders of magnitude above the 50 GHz drive. We therefore do not claim a microscopic realization of $U \approx 0.05$–$0.40$ in a uniformly magnetized Co:YIG cell of this size. $U$ is used below as a normalized model parameter, and the results are statements about the normalized dynamics at a given $U$, not about a specific microscopic realization; identifying a physical system that reaches that range — a defect-localised mode, a magnetic nanocontact, or a molecular magnet — is left to future work. The numerical results below should therefore be read as a study of the normalized model rather than as a simulation of a specific Co:YIG mode.

Microscopic origin of $U$ and its dependence on cell size. The Kerr coefficient is not a free material parameter: it is fixed by the spin content of the mode. The uniform mode of a saturated ferrimagnet is described by a single macrospin, $S_z = M_z V/(\gamma\hbar)$, reduced by one unit per magnon through the Holstein–Primakoff substitution $S_z = N_{eff} - n$. The effective spin content of the mode is therefore its net angular momentum, $N_{eff} = M_s V/(g\mu_B)$, and not the sum of the bare ion spin lengths: in a ferrimagnet the two $Fe^{3+}$ sublattices are antiparallel, and only the uncompensated moment participates in the acoustic branch. The mode saturates after $2N_{eff}$ magnons, so the anharmonic shift per magnon is $\omega_a/(2N_{eff})$ and $U = \omega_a/(2N_{eff}\Omega)$. This normalization is the one already implied by $\omega_a = \gamma(2K_u/M_s)$ used above, which is the macrospin anisotropy frequency. It can be checked without reference to any spin count at all: substituting $N_{eff} = M_s V/(g\mu_B)$ into $\omega_a/(2N_{eff})$ gives a Kerr coefficient $K = \hbar\gamma^2 K_u/(M_s^2 V)$, which depends only on measurable quantities and is the standard cavity-magnonics result for the anisotropy-induced Kerr term of the Kittel mode. For the $20 \times 20 \times 10$ nm³ cell, $M_s = 1.4 \times 10^5$ A m$^{-1}$ gives a net spin content $N_{eff} = 3.0 \times 10^4$, a per-magnon Kerr scale $K/2\pi = 3.3$ MHz and $U = 6.6 \times 10^{-5}$ at $\Omega/2\pi = 50$ GHz. Part of the cavity-magnonics literature instead normalizes the macrospin by the total ion spin density, which for yttrium iron garnet at 310 K gives an $N_{eff}$ larger by a factor of seven and a $U$ smaller by the same factor; every conclusion drawn below from the smallness of $U$ is therefore conservative with respect to that alternative, and every conclusion drawn from the criterion of Section 8 is the less restrictive of the two. Since $N_{eff}$ is proportional to the volume, $U$ scales as $1/V$: the cell size is the only lever, and it acts in the favourable direction for the switching energy at the same time, because at a fixed absorbed-energy density of 6 J cm$^{-3}$ the energy per event also scales as $V$. Shrinking the cell from 20 × 20 × 10 nm³ to a 10 nm cube raises $U$ to $2.7 \times 10^{-4}$ and lowers the switching energy from 24 aJ to 6 aJ; a 5 nm cube gives $U = 2.1 \times 10^{-3}$ at 0.75 aJ. The lever runs out at the thermal-stability limit: requiring $K_u V \geq 40 k_B T$ at 310 K gives $V \geq 342$ nm³, a 7 nm cube, where $U = 7.7 \times 10^{-4}$. Even discarding thermal stability altogether, a single YIG cubic unit cell (1.90 nm³, net spin content 14.3) reaches only $U = 0.14$. Reaching $U = 0.05$ requires $N_{eff} = 40$, about sixteen formula units; $U = 0.1$ requires $N_{eff} = 20$, about one cubic unit cell; $U = 0.40$ requires $N_{eff} = 5$, two formula units. That is the regime of a few-spin quantum object, not of a magnon mode, and it is why no choice of geometry within the magnonic setting closes the gap.

To examine the model's behaviour we ran a 100-level Fock-basis calculation in a frequency, pulse-duration and damping context motivated by experiment on Co-doped garnet films (Fig. 2). Taking the ultrafast-switching benchmark of Stupakiewicz et al. [18], a 50-fs excitation pulse, together with the 50 GHz operating frequency adopted throughout, the expected bosonic spectral divergence does not occur. In this section $U$ is treated as a free normalized parameter of the single-mode model; it is not identified with the anisotropy of the $Co^{2+}$ ions, since the criterion derived in Section 8 fixes it as $U = \omega_a/(2N_{eff}\Omega)$, that is by the anisotropy field together with the spin content of the mode.

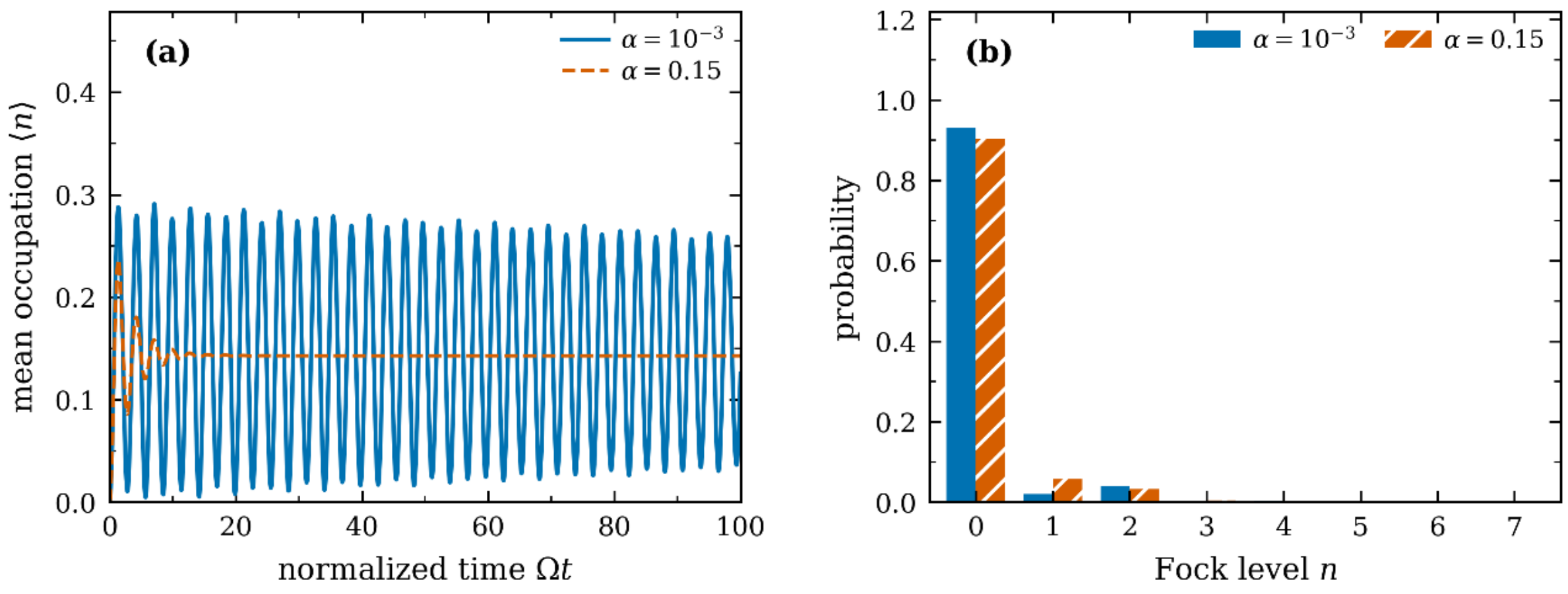

Figure 1. Long-time numerically bounded dynamics and Fock-space distribution. Lindblad evolution of the driven magnon mode ($U = 0.1$, $\eta = 1.2$, $\kappa = 2\alpha\Omega$, 40-level basis, vacuum initial state, $\Omega t = 100$) at two damping levels. (Left) Mean occupancy $\langle n \rangle$: both traces are numerically bounded; the low-damping trace ($\alpha = 10^{-3}$) shows persistent oscillations with $\max_t \langle n \rangle = 0.2909$ and no secular growth. (Right) Final Fock-level probabilities: $P_0 + P_1 > 0.95$ for both cases and $P(n \geq 10) < 3\times10^{-7}$.

The Lindblad dynamics shown in Figs. 1–5 were integrated with the adaptive master-equation solver implemented in QuTiP. The damping rate was set to $\kappa = 2\alpha\Omega$, so that the mean-field mode amplitude decays at the Gilbert rate $\alpha\Omega$. Numerical convergence was verified with respect to the Fock-space truncation, the solver tolerances and the temporal output sampling. The last of these is not a formality: the solver integrates adaptively between the requested output times but returns expectation values only at those times, so a coarse output grid systematically misses the narrow maxima of $\langle n \rangle$. This is not a hypothetical concern: for the $U = 0.1$, $\alpha = 10^{-3}$ protocol the 100-cycle run of Fig. 2 samples at 20 points per drive cycle and records a maximum of 0.2843, whereas the shorter run of Fig. 1, sampled at about 126 points per cycle, records 0.2909, and a dedicated refinement to 1000 points per cycle gives 0.2909. The converged value of approximately 0.291 is the one underlying the bounds quoted for that protocol below. Selected parameter sets were independently cross-checked by direct propagation of the time-independent Liouvillian corresponding to Eq. (5), yielding consistent final-state populations. Throughout the evolution we monitored the boundedness of the mode occupation and the energy balance of the system.

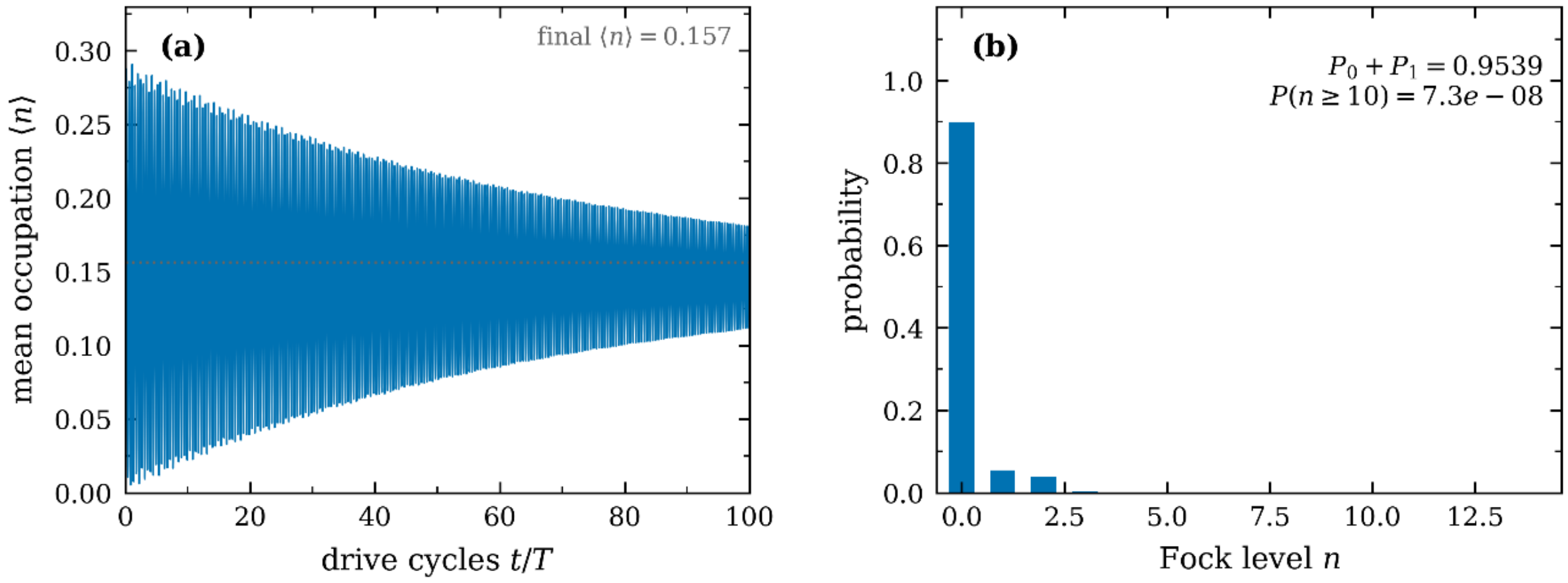

Figure 2. Long-time stability over 100 operational cycles (100-level basis). Same protocol as Fig. 1 at $\alpha = 10^{-3}$, $\Omega/2\pi = 50$ GHz. (Left) $\langle n \rangle$ vs cycle count: numerically bounded with final value 0.157, no secular growth. Note: the recorded maximum 0.2843 is sampling-resolution limited (20 pts/cycle); the converged value is 0.2909 (see text). (Right) Final Fock distribution: $P_0+P_1 = 0.9539$, $P(n \geq 10) = 7.3\times10^{-8}$, and no measurable population above $n = 90$.

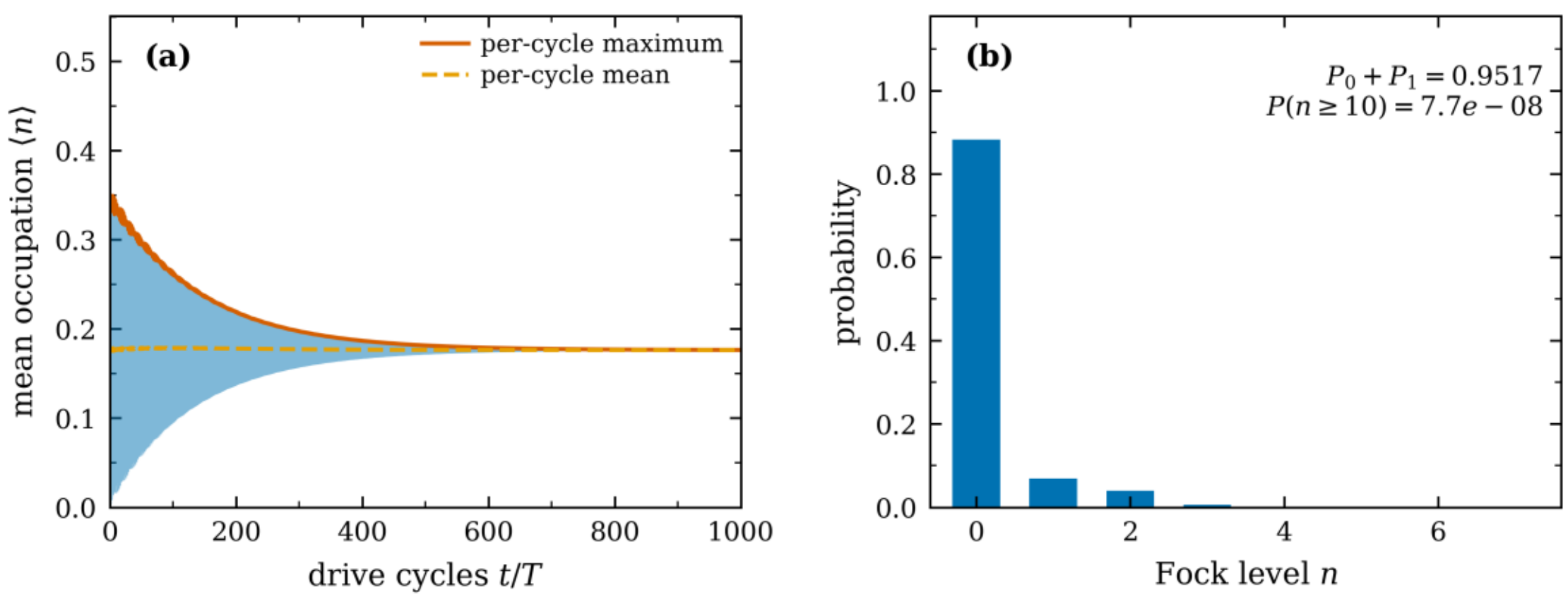

Figure 3. Full quantum verification over 1000 drive cycles (35-level basis, converged). (Left) $\langle n \rangle$ vs cycle: per-cycle max (red) falls from 0.349 to 0.177 and per-cycle mean (orange) plateaus at 0.176; $\langle n \rangle$ = 0.183, 0.177, 0.176 at cycles 100, 500, 900. (Right) Final Fock distribution at cycle 1000: $P_0$+$P_1$ = 0.9517, $P(n \geq 10) = 7.7\times10^{-8}$.

The relative integration tolerance was set to $10^{-7}$. Convergence with respect to the Fock-space truncation, the solver tolerances and the temporal output sampling is reported separately above and in Supplement 1; the numerical accuracy is a property of the solver settings and the basis cutoff, not of the nonlinear term. What the nonlinearity does is dynamical: it confines the driven state to a bounded low-occupation region of the Fock space, which is what keeps the truncated basis adequate.

The experimentally reported absorbed-energy density of approximately 6 J/cm³ [18] corresponds to a switching-event energy of approximately 22–24 aJ for a 20×20×10 nm³ cell (Section 3). This conversion establishes the energy budget of the switching event, but does not by itself determine the magnetic anisotropy constant $K_{eff}$ or the Kerr coefficient $U$. In the present numerical model, $U$ is a normalized nonlinear-frequency-shift parameter whose value is chosen on the basis of the dynamical stability and low-occupancy behaviour surveyed in Fig. 4, not derived from the absorbed-energy density above. That mapping is carried out in Section 5.1, from the anisotropy constant, the saturation magnetization and the mode volume, and gives $U = \omega_a/(2N_{eff}\Omega)$ with $N_{eff} = M_s V/(g\mu_B)$.

To account for open-system dynamics, we incorporate a dissipation term $\gamma$ into our 100-level Fock basis model, consistent with our previous theoretical framework [4]. Explicitly, the open-system dynamics are governed by the Lindblad master equation for the density matrix $\rho$, with $\hat{H}(t)$ given by Eq. (4) and a single dissipation channel on the mode operator $\hat{a}$, at rate $\kappa$=$2\alpha\Omega(t)$ chosen so that the resulting mean-field amplitude decay matches the Gilbert-damping rate $\alpha\Omega(t)$ used throughout this work. The exact Heisenberg-Langevin reduction of this master equation, and the mean-field (Bogoliubov) closure used to obtain the semiclassical $u(t)$, $v(t)$ equations of motion used in Sections 5.2–5.3, are given in Supplement 2. The system dynamics is described by the density operator $\rho$, whose time evolution is governed by the Lindblad master equation

$$d\rho/dt = -(i/\hbar)[\hat{H}(t), \rho] + \kappa\,(\hat{a}\rho\hat{a}^{\dagger} - \tfrac{1}{2}\{\hat{a}^{\dagger}\hat{a}, \rho\}\,),\ \ \kappa = 2\alpha\Omega(t) \quad (5)$$

Finite-temperature magnon injection is neglected in this section; it is reinstated by the explicit thermal operators of Section 6. For the Co-doped garnet system, the dissipation rate $\gamma$ is parameterized via the Gilbert damping constant $\alpha$, where $\gamma = \alpha\Omega$. Using the experimentally derived $\alpha \approx 0.15$ [20], we note that dissipation here removes excitation from the already nonlinearly bounded manifold; it is not the origin of the confinement, which is set by the nonlinear detuning and persists at low damping (Fig. 1). With $\Omega = 2\pi \times 50$ GHz this gives a Gilbert damping rate $\alpha\Omega \approx 4.7 \times 10^{10}$ s$^{-1}$, and a Lindblad rate $\kappa = 2\alpha\Omega \approx 9.4 \times 10^{10}$ s$^{-1}$. Numerical verification used the parameters specified in the Supplement 1: a non-adiabaticity factor $\eta = 1.2$ and a nonlinear regulator strength $U_{\mathrm{audit}} = 0.12$ (normalized). $U_{\mathrm{audit}} \approx 0.12$ is the analytically estimated stabilization crossover, while $U_{\mathrm{audit}} \approx 0.55$ together with $\alpha \approx 0.01$ is the representative operating point used below (see Section 5). In the convention of Eq. (4) these correspond to $U = 0.06$ and $U = 0.275$, that is to $\xi = \eta/U = 20$ at the analytic crossover and $\xi \approx 4.4$ at the representative operating point.

Dependence of the confined state on the nonlinear regulator. We map the confined state as a function of the nonlinear coefficient by integrating Eq. (5) on a grid of regulator strengths from $U_{\mathrm{audit}} = 0$ to $U_{\mathrm{audit}} = 0.8$ at $\eta = 1.2$ and $\alpha = 0.01$, using the Kerr term $(U_{\mathrm{audit}}\Omega/2)\hat{a}^{\dagger 2}\hat{a}^2$ that is also used for the resonant calculations of Section 5.2. That normalization carries half the anharmonicity of the convention of Eq. (4), so these runs are labelled by $U_{\mathrm{audit}}$ throughout and are converted by $U = U_{\mathrm{audit}}/2$ wherever they are compared with the criterion of Section 8. The same holds for almost every other calculation quoted here, and the exceptions are worth listing once. Figs. 4–6, the protocol audit, the five-phase and phase-step protocols, the semiclassical stress test and the 310 K frequency scan of Fig. 10 are all integrated with $(U_{\mathrm{audit}}\Omega/2)\hat{a}^{\dagger 2}\hat{a}^2$, so every numerical value of the nonlinear coefficient quoted for them is a $U_{\mathrm{audit}}$ value. The 1000-cycle calculation of Fig. 3 uses $U\Omega\hat{a}^{\dagger 2}\hat{a}^2$ directly, so its $U = 0.1$ is already an Eq. (4) value. Figs. 1 and 2 use the earlier $U\hat{n}^2$ form, which differs from both by an additional linear shift $U\Omega\hat{n}$; they are retained because they establish boundedness rather than a numerical value. The conversion is not a matter of bookkeeping. The mean-field balance of Section 5.3 predicts a saturated occupation $\eta/(4U)$; at the representative resonant point ($\eta = 1.2$, $U_{\text{audit}} = 0.55$, hence $U = 0.275$) this gives 1.09 against the 1.213 returned by the full Lindblad calculation, an agreement to 11%, whereas reading $U_{\text{audit}}$ directly as $U$ would predict 0.545 and miss by a factor of two. The numerics therefore select the conversion independently of the algebra. Because the driven state oscillates, we report both the maximum mean occupancy $\max_t\langle n\rangle$ reached at any time and the mean occupancy $\langle n\rangle$ averaged over the second half of the run, together with the corresponding late-time populations of the low-lying manifold (Fig. 4). The outcome is monotonic rather than structured: increasing $U_{\text{audit}}$ lowers $\max_t\langle n\rangle$ from 0.551 at $U_{\text{audit}} = 0$ to 0.185 at $U_{\text{audit}} = 0.8$, and the late-time mean occupancy from 0.280 to 0.094 over the same range, with no local optimum, threshold or kink anywhere in the scanned interval.

Two consequences follow from Fig. 4 and are stated explicitly. First, $U_{\text{audit}} \approx 0.55$ is a representative rather than a preferred value. It lies well inside the strongly confined nonlinear regime, and neighbouring values of the reference parameter set behave comparably: the late-time two-level population $P_0 + P_1$ is 0.9645, 0.9684, 0.9728 and 0.9764 at $U_{\text{audit}} = 0.50$, 0.55, 0.60 and 0.80, so $U_{\text{audit}} = 0.55$ is one of a range of comparable choices rather than an optimum fixed by the simulation. At this representative point the simulation gives $\max_t\langle n\rangle = 0.228$, a late-time mean occupancy $\langle n\rangle = 0.117$, a late-time $P_0 + P_1 = 0.9684$, $P(n \geq 2) = 0.0316$ and $P(n \geq 10)$ below $10^{-9}$. Second, $U_{\text{audit}} \approx 0.12$ marks an analytic crossover rather than a numerically observed threshold: between $U_{\text{audit}} = 0.10$ and 0.15 the maximum occupancy changes smoothly (0.413 → 0.396 → 0.374), the curve and its slope remaining continuous throughout. We therefore refer to $U_{\text{audit}} \approx 0.12$ throughout as the analytically estimated stabilization crossover of Section 2, marked in Fig. 4 for reference, rather than as a numerically resolved stability threshold.

The same scan also fixes the correct description of the confined state. At the representative point the late-time mean occupancy is $\langle n\rangle \approx 0.12$, so almost the entire probability weight sits in the vacuum: a high value of $P_0 + P_1$ does not by itself imply that one magnon is present. Resolving the two lowest populations individually at $U_{\text{audit}} = 0.55$ gives, at the final time, $P_0 = 0.9173$ and $P_1 = 0.0511$, so the confined state under this static protocol is a low-excitation state within the vacuum–single-magnon manifold rather than a state containing one magnon on average, and we use that description throughout. The resonantly modulated protocol of Section 5.2 places the same nonlinear mode in a different, order-unity occupation regime; the two protocols are compared directly there. A systematic transient scan may quantify the largest dissipatively generated $P_1$ attainable within the pair-driven model, and would usefully be accompanied by calculations of the purity of the final state and of the second-order correlation function; such a scan would not, however, demonstrate coherent one-magnon preparation, because the Hamiltonian contains no parity-breaking single-quantum drive (Section 5.2).

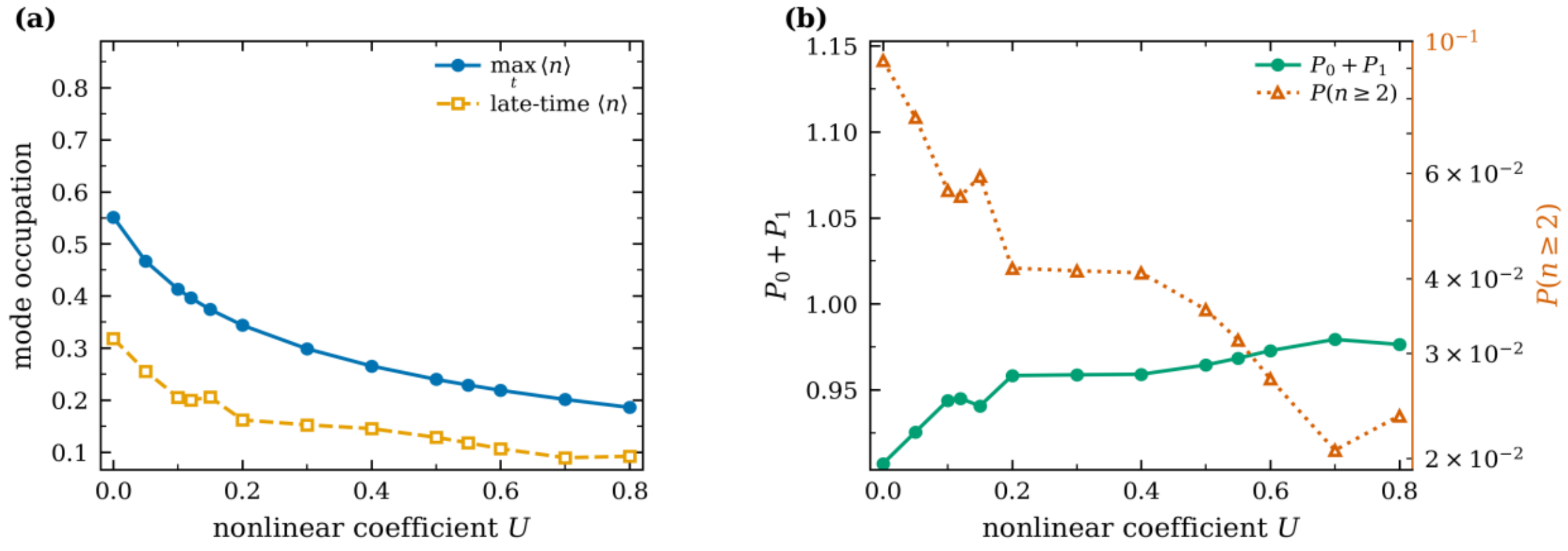

Figure 4. Confined occupancy versus nonlinear coefficient $U$ (static effective pair drive, Kerr term $(U_{\text{audit}}\Omega/2)\hat{a}^{\dagger 2}\hat{a}^2$, 50-level Lindblad, $\eta = 1.2$, $\alpha = 0.01$). (a) $\max_t\langle n\rangle$ and late-time mean $\langle n\rangle$ vs $U$: both decrease monotonically with no threshold. (b) Late-time $P_0+P_1$ and $P(n \geq 2)$ vs $U$: confinement improves smoothly apart from small non-monotonic wiggles at the 0.5% level, with no discontinuity or sharp threshold anywhere in the scanned interval. Dashed line: analytic crossover $U_{\text{audit}} = 0.12$. Dotted line: representative operating point $U_{\text{audit}} = 0.55$.

The switching-event energy corresponding to this cell (20×20×10 nm³ at the experimental energy density of 6 J/cm³ [18]) is approximately 22–24 aJ; as stated above, this is an energy context for the calculation, not a quantity derived from it.

The results (Fig. 2) show that the mean occupancy $\langle n\rangle$ remains bounded below 0.30 throughout this full 100-level Fock-basis (Lindblad) simulation, with the population concentrated in the lowest levels and a population in levels $n \geq 10$ of $7.3 \times 10^{-8}$, indicating that $U = 0.1$ used in this particular Lindblad simulation suppresses unbounded Fock-space occupation for the protocol investigated. This calculation does not by itself establish $U = 0.1$ as a stability threshold, nor as a preferred operating point; the values used elsewhere in the paper are the analytically estimated crossover $U_{\text{audit}} \approx 0.12$ and the representative operating point ($U_{\text{audit}} \approx 0.55$, $\alpha \approx 0.01$), both examined in Fig. 4. Earlier comparisons performed with non-identical drive implementations suggested a quantitative discrepancy between the reduced coupled-mode and full Lindblad descriptions. That comparison has since been repeated under matched conditions and is reported in Section 5.2.

### 5.2. Bogoliubov–Lindblad bridge and pair-phase control

Two complementary representations are used in this work. The first is a reduced coupled-mode description in terms of the Bogoliubov amplitudes $u(t)$ and $v(t)$. For a quadratic bosonic evolution initialized in the vacuum, the negative-frequency amplitude determines the mean mode occupation, $n_B(t) = |v(t)|^2$. The second is the full open-system evolution of the density matrix $\rho(t)$ in a truncated Fock basis, from which the mean occupation and the level populations are obtained as $n_L(t) = \mathrm{Tr}[\rho(t)\, \hat{a}^\dagger \hat{a}]$ and $P_j(t) = \langle j|\rho(t)|j\rangle$. These quantities must not be conflated: $n_B$ and $n_L$ characterize the mean number of excitations, whereas $P_1$ is the probability of finding exactly one excitation. A mean occupation close to unity therefore does not imply a pure single-magnon state.

Both representations are driven by the same externally prescribed modulation protocol, and both use the Kerr term $(U_{\mathrm{audit}}\Omega/2)\hat{a}^{\dagger 2}\hat{a}^2$, so that $U_{\mathrm{audit}}$ has the same meaning here as in the confinement scan of Fig. 4. The comparison reported in this subsection uses $\hat{H}(t) = \Omega\hat{a}^\dagger\hat{a} + (U_{\mathrm{audit}}\Omega/2)\hat{a}^{\dagger 2}\hat{a}^2 + (\mathcal{G}(t)/2)(\hat{a}^{\dagger 2} + \hat{a}^2)$ with $\mathcal{G}(t) = \eta\Omega \sin(2\Omega t)$, that is, a pair coefficient modulated at twice the mode frequency. This differs from the confinement scan of Fig. 4 only in the drive: there the pair coupling is the time-independent term $\eta\Omega/4$, a static effective pair-drive Hamiltonian in the frame used for that calculation, whose relation to a laboratory-frame modulation is not derived here. Both protocols give the same parametric growth rate $h = \eta\Omega/2$. The two therefore share the nonlinear term and the parameters $U_{\mathrm{audit}}$, $\eta$ and $\alpha$, and differ only in whether the pair drive is static or resonantly modulated. That single difference is enough to move the system between two bounded but very different regimes: run under the same code with identical $U_{\mathrm{audit}} = 0.55$, $\eta = 1.2$ and $\alpha = 0.01$, the static protocol gives a late-time mean occupation of 0.117 ($P_0 + P_1 = 0.968$), whereas the resonantly modulated protocol gives 1.213 ($P_0 + P_1 = 0.645$) — a factor of about ten in mean occupation. We state this explicitly because the shared parameter labels would otherwise suggest a contradiction between Figs. 3 and 4 where there is none.

To test whether the reduced description identifies the same occupation scale as the full quantum model, we integrated both sets of equations using a matched drive protocol, nominal nonlinear scale, damping convention and time interval, with the operator Kerr interaction represented in the reduced model by a mean-field frequency shift $\Delta = U_{\mathrm{audit}}\Omega|v|^2$, which is the Hartree closure of the audit Hamiltonian and equals $2U\Omega|v|^2$ in the convention of Eq. (4). For $U_{\mathrm{audit}} = 0.55$ (that is, $U = 0.275$), $\eta = 1.2$, $\alpha = 0.01$ and $\kappa = 2\alpha\Omega$, the late-time coupled-mode average is $|v|^2 = 1.1165$ (standard deviation 0.4940), whereas the full Lindblad calculation gives $\langle n\rangle = 1.2133$ (standard deviation 0.2224). The ratio of the two averages is 0.920, so the reduced model identifies the late-time mean occupation scale to within approximately 8% for this parameter set. The agreement concerns the time-averaged occupation only. The instantaneous trajectories differ appreciably — the maximum deviation $|n_B - n_L|$ is 2.56 and the root-mean-square deviation is 0.72, i.e. comparable to the mean itself — and the coupled-mode closure replaces the operator Kerr term by a single mean frequency shift and does not retain the Fock-state distribution or the higher-order quantum correlations (Fig. 5).

The anomalous correlator requires a consistent Bogoliubov convention: for the transformation used in the reduced equations the quantity corresponding to the Lindblad correlator $\langle \hat{a}\hat{a} \rangle$ is $uv^*$, not $uv$. With that convention the reduced model tracks the pair phase considerably better than the pair amplitude — the mean phase offset is 0.015 rad against a root-mean-square magnitude difference of 0.482 — and the phase of the correlator rotates at twice the mode frequency, with a residual variation of 0.224 rad (about 13°) once the carrier rotation is removed. We report the latter as a preliminary characterization of the ensemble-averaged pair-phase stability and not as a phase-diffusion rate. The full convention audit is given in Supplement 4.

The Lindblad populations clarify what this order-unity occupation regime actually is. At the final time the distribution is $P_0 = 0.3167$, $P_1 = 0.3183$ and $P_2 = 0.2208$, with $P(n \geq 2) = 0.3650$ and $P(n \geq 10) \approx 3.8 \times 10^{-6}$. At the time at which $P_1$ reaches its maximum value of 0.3817, the mean occupation is $\langle n \rangle = 1.0021$, while $P_0 = 0.3389$ and $P(n \geq 2) = 0.2793$, so substantial probability remains in both the vacuum and the higher Fock levels. The regime identified by the coupled-mode calibration should therefore be described as a bounded order-unity mean-occupation regime distributed over the low-lying Fock manifold, and not as the preparation of a pure single-magnon Fock state.

This limitation is structural rather than numerical. The Hamiltonian above contains only the diagonal terms $\hat{a}^\dagger\hat{a}$ and $\hat{a}^{\dagger 2}\hat{a}^2$ and the pair term $\hat{a}^{\dagger 2} + \hat{a}^2$, all of which commute with the parity operator $(-1)^{\hat{n}}$. Starting from the vacuum, the unitary part of the evolution therefore populates even Fock levels only, and the odd populations — including $P_1$ — are generated only indirectly, through single-quantum decay of the pair-excited even states via the loss channel of the Lindblad term. This is directly visible in the computed parity expectation value: $\langle (-1)^{\hat{n}} \rangle$ retains the value 0.8935 at the end of the static run, whereas the resonant drive brings it down to 0.1543, showing that strong pair creation combined with single-quantum loss efficiently mixes the even and odd sectors. Because the Hamiltonian contains no coherent single-quantum coupling, the present mechanism does not provide deterministic coherent preparation of $|1\rangle$, and the calculated $P_1$ should be interpreted as a dissipatively generated transient population rather than as a Fock-state preparation fidelity. At the resonantly driven representative point its maximum is $P_1^{max} = 0.3817$, and the associated state remains broadly distributed over the vacuum and the higher Fock levels. Coherent preparation of a one-magnon state would require a parity-breaking, single-quantum drive term proportional to $\hat{a} + \hat{a}^\dagger$, together with a Kerr blockade of the $0 \rightarrow 2$ transition; such a drive is outside the model studied here. Establishing a rigorous upper bound on $P_1$ for the driven-dissipative model would require a separate analytical treatment, which we do not attempt here.

The roles of the two descriptions are thus complementary rather than contradictory. The coupled-mode equations provide a computationally efficient representation of nonlinear saturation and phase evolution over broad parameter ranges, whereas the Lindblad treatment determines the quantum number statistics, the leakage and the mixed-state structure. Agreement of their late-time mean occupations supports the physical reality of the bounded occupation scale identified by the coupled-mode calibration at ($U_{audit} \approx 0.55$, $\alpha \approx 0.01$), while the Fock-resolved calculation sets the limits of its single-magnon interpretation. Accordingly, the present results support collective-mode amplitude and phase encoding within a bounded low-occupation manifold; they do not establish deterministic preparation of the one-magnon Fock state.

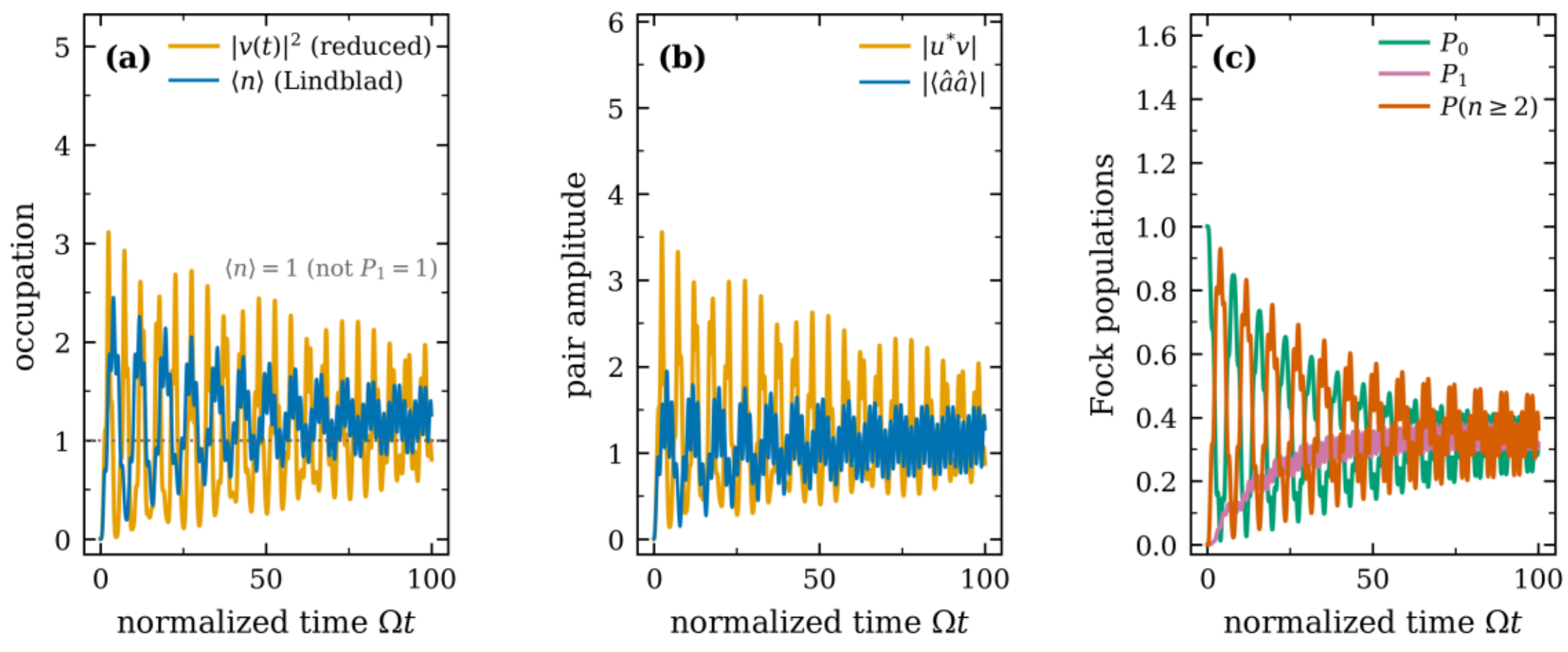

Figure 5. Bogoliubov–Lindblad bridge. Mean occupation $n_B(t) = |v(t)|^2$ (reduced Bogoliubov, orange) vs $n_L(t) = \mathrm{Tr}[\rho\, \hat{a}^\dagger \hat{a}]$ (full Lindblad, blue) for $U_{audit} = 0.55$, $\eta = 1.2$, $\alpha = 0.01$, resonant drive, $\Omega t = 100$. Late-time means: $n_B = 1.117$, $n_L = 1.213$ (agreement ~8%); instantaneous deviations are large (RMS = 0.72) because the reduced model lacks the full Fock statistics. Dashed line: $\langle n\rangle = 1$ (not $P_1 = 1$).

The phase of the anomalous correlator can also be addressed directly through the phase of the drive. Repeating the resonant calculation with $G(t) = \eta\Omega \sin(2\Omega t + \varphi_d)$ for $\varphi_d = 0, \pi/4, \pi/2, 3\pi/4$ and $\pi$ gives a rotating-frame pair phase that follows the drive phase with a fitted slope of −0.99984 ($R^2 = 0.9999993$, residual $9.3 \times 10^{-4}$ rad), while the late-time mean occupation varies by only about 1% and the pair-correlation modulus $|\langle\hat{a}\hat{a}\rangle|$ by about 2.5% across the full range of $\varphi_d$ (Fig. 6). Two points should be stated plainly about how this result is to be read. First, for a strictly periodic drive a shift of $\varphi_d$ is equivalent to a shift of the time origin, so at late times the relation $\Phi_{pair} = \Phi_0 - \varphi_d$ follows from time-translation symmetry; the measured slope is therefore a check that the pair phase is sharp and that the calculation is internally consistent, not an independent demonstration of dynamical controllability. Second, the same symmetry implies that the late-time occupation should be exactly independent of $\varphi_d$; the residual 1–2.5% variation reflects the memory of the vacuum initial condition, which at $\Omega t = 100$ has decayed by only one e-folding of the damping time $(\alpha\Omega)^{-1} = 100$, and sets the numerical accuracy of this test rather than a physical dependence.

The dynamical content is instead supplied by a separate step-response calculation, in which $\varphi_d$ is held at zero and then switched abruptly at $\Omega t = 100$, so that the protocol is no longer periodic and the time-translation argument no longer applies. Steps of $\pi/8$, $\pi/4$, $\pi/2$ and $\pi$ were applied together with a null control, and the response was extracted from the rotating-frame complex correlator $\langle\hat{a}\hat{a}\rangle\exp(2i\Omega t)$ rather than from its argument, so that no phase unwrapping is needed when the amplitude becomes small. Three conclusions follow. The steady-state phase transfer is essentially exact for every step size, $-\Delta\Phi/\varphi_{step} = 0.995$–$1.000$. The complex correlator relaxes to its new steady value in $\tau_{resp} \approx 15$–$26\ \Omega^{-1}$, that is five to eight drive periods and well below the damping time $(\alpha\Omega)^{-1} = 100\ \Omega^{-1}$, so the phase is reset by the drive and not by dissipation; we regard this figure as provisional, because at $\Omega t = 100$ the pre-step state has itself relaxed for only one damping time. And the phase-only description has a finite operating range: for steps up to $\pi/4$ the pair modulus stays at 45–70% of its steady value, whereas at $\pi/2$ and $\pi$ it is driven below 2% and the mean occupation more than doubles transiently. The full set of response times, modulus minima and occupation excursions, together with the null-control subtraction, is tabulated in Supplement 5.

In the calculated protocols the trajectories associated with the larger phase steps pass close to the origin of the complex-correlation plane, where the modulus is small and the instantaneous phase is ill-defined, whereas the $\pi/8$ and $\pi/4$ trajectories do not. We note, without claiming it as a derived result, that this is what would be expected if a reversal of the pump phase leaves the existing pair coherence anti-aligned with the new drive, so that it is damped before being rebuilt at the new phase rather than rotated into it. The behaviour is therefore graded: a $\pi/8$ step is a clean phase rotation, a $\pi/4$ step already constitutes coupled amplitude–phase switching, and $\pi/2$ and $\pi$ cannot be described as phase-only switching at all. This is a design constraint rather than a failure of the mechanism, and it bears directly on collective-mode encoding: a binary code based on pair phases separated by $\pi$ cannot be switched in this model without collapsing and rebuilding the pair amplitude. The step-response traces and the complex-plane trajectories are given as Figure S3 in the Supplementary Material.

**The transfer coefficient is kinematic.** Unlike the Fock-manifold confinement, which requires $U$ to be large enough that the saturated occupation $\eta/(4U)$ stays below one, the phase transfer does not depend on the Kerr coefficient at all. Repeating the five-phase protocol at $U_{\text{audit}} = 0.55$, 0.10 and 0.02 — that is $U = 0.275$, 0.05 and 0.01 — over which the late-time occupancy rises from 1.2 to 30 magnons and the system leaves the low-occupancy manifold entirely, leaves the slope at −1.00000 with $R^2 = 0.999999$ in every case; the within-run circular standard deviation even improves slightly, from 13.0° to 10.1°, as the larger occupancy reduces the relative phase noise. The pair phase is locked to the pump phase by the structure of the parametric coupling rather than by the nonlinear confinement, so the pair-phase observable survives in exactly the parameter range where Fock-state confinement does not. This matters for the assessment of Section 8: the mean-field screening rule $N_{\text{eff}} < 1/\alpha$ applies to the confined-regime estimate, whereas phase readout needs only that the parametric drive exceed the damping.

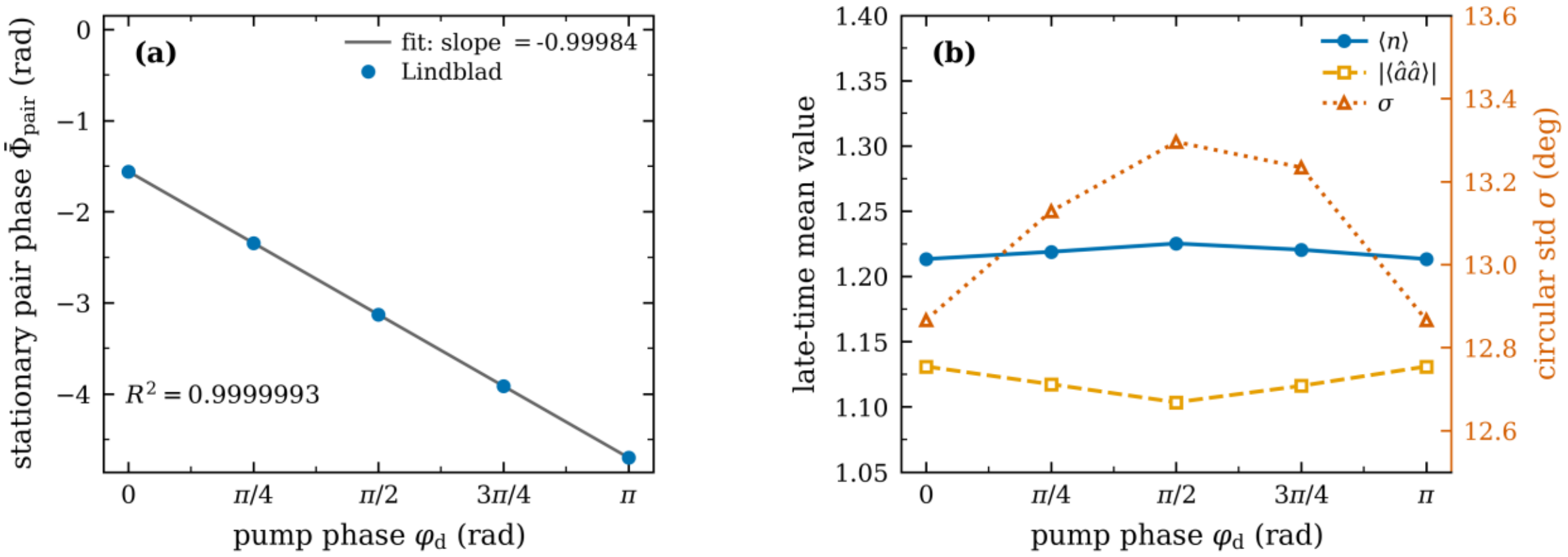

Figure 6. Pair-phase transfer. Full Lindblad ($U_{\text{audit}} = 0.55$, $\eta = 1.2$, $\alpha = 0.01$, $N = 50$, resonant drive) for five pump phases $\varphi_d$. (a) Late-time rotating-frame pair phase $\Phi_{\text{pair}}$: slope = −0.99984, $R^2 = 0.9999993$, residual $9.3\times10^{-4}$ rad. Error bars show the within-run circular std (~13°), not inter-sample uncertainty. (b) $\langle n\rangle$ and $|\langle\hat{a}\hat{a}\rangle|$ are insensitive to $\varphi_d$ (1–2.5%).

### 5.3. Semiclassical parameter survey

The Fock-resolved Lindblad results above fix the confinement quantitatively at specific parameters. The reduced coupled-mode (Bogoliubov) description of Section 5.2, being far cheaper, lets us survey the same confinement across a much wider range of the Kerr coefficient $U$ and the drive intensity $\eta$; it captures the mean-occupation scale, though not the Fock-state statistics (Section 5.2). We use it here only for that broad-range survey (method in Supplement 2); the outcome of that survey is shown in Fig. 7.

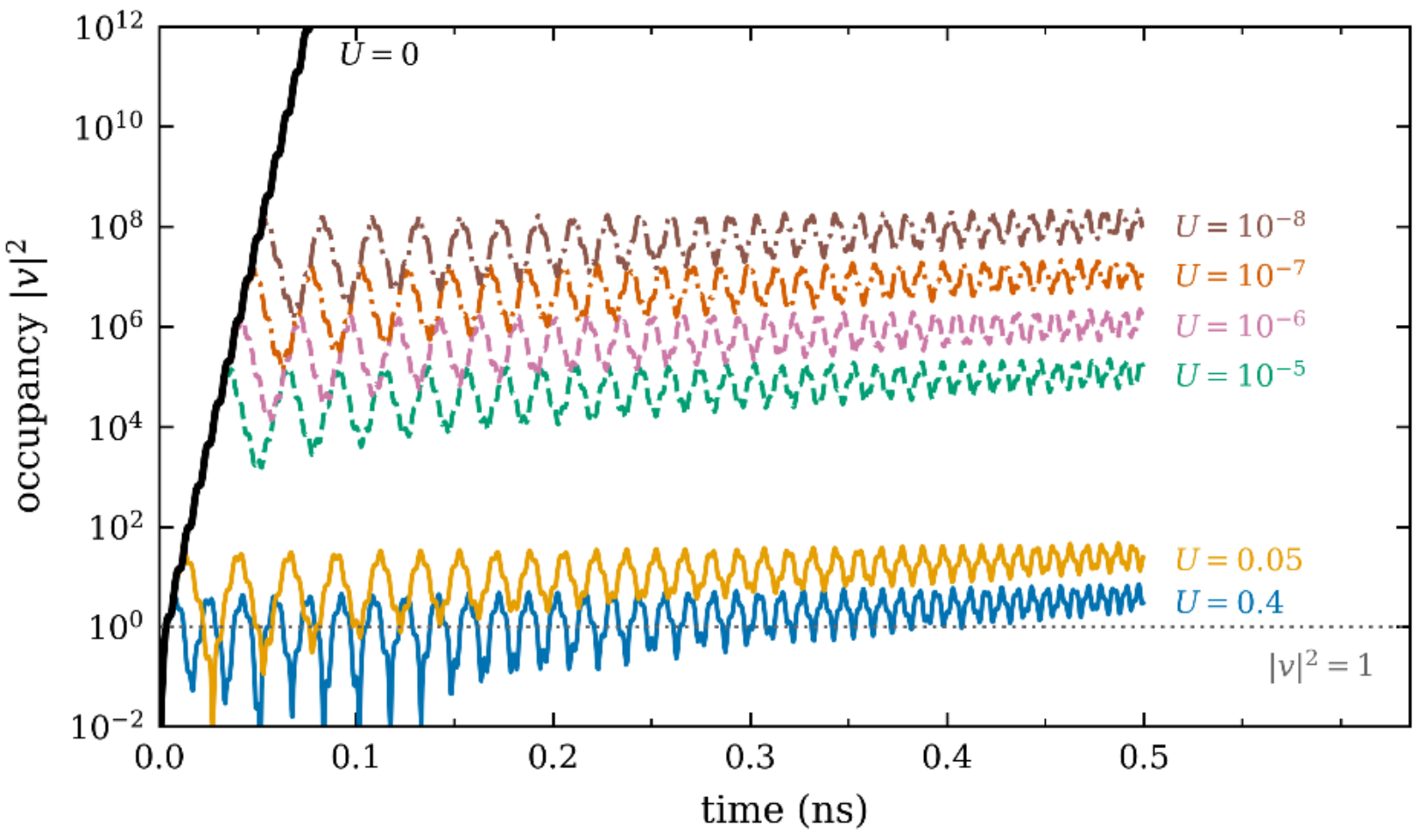

Figure 7. Numerical verification of the non-adiabatic parametric excitation stability and the nonlinear regulator mechanism. The black curve ($U = 0$) leaves the top of the plotted range at about 0.08 ns and continues to grow without bound; the vertical axis is truncated at $10^{12}$ so that the stabilized curves remain resolvable.

Figure 7 follows the magnon occupancy $|v|^2$ in time under a non-adiabatic drive at $\eta = 1.2$. Without the Kerr term ($U = 0$, black curve) the occupancy diverges monotonically and exponentially, an unconstrained and unstabilized regime in which information is lost to the parametric excitation. It leaves the plotted range well before the end of the interval, standing some seventy orders of magnitude above the top of the axis at 0.5 ns, so the truncation is in the axis and not in the calculation. Switching on the Kerr nonlinearity brings dynamic localization. With anharmonicity as small as $U = 10^{-8}$ to $10^{-5}$ the occupancy does level off, but at $10^4$–$10^8$, which is of no use for coherent operation. At the operational values $U = 0.05$ to 0.4 (orange and blue curves) the nonlinearity detunes the mode away from parametric resonance and locks the occupancy inside a bounded low-occupancy regime, suppressing higher-order excitation. What remains is a strongly confined sector dominated by the two lowest Fock levels, a bounded regime worth examining further as a low-energy resonant operating sector.

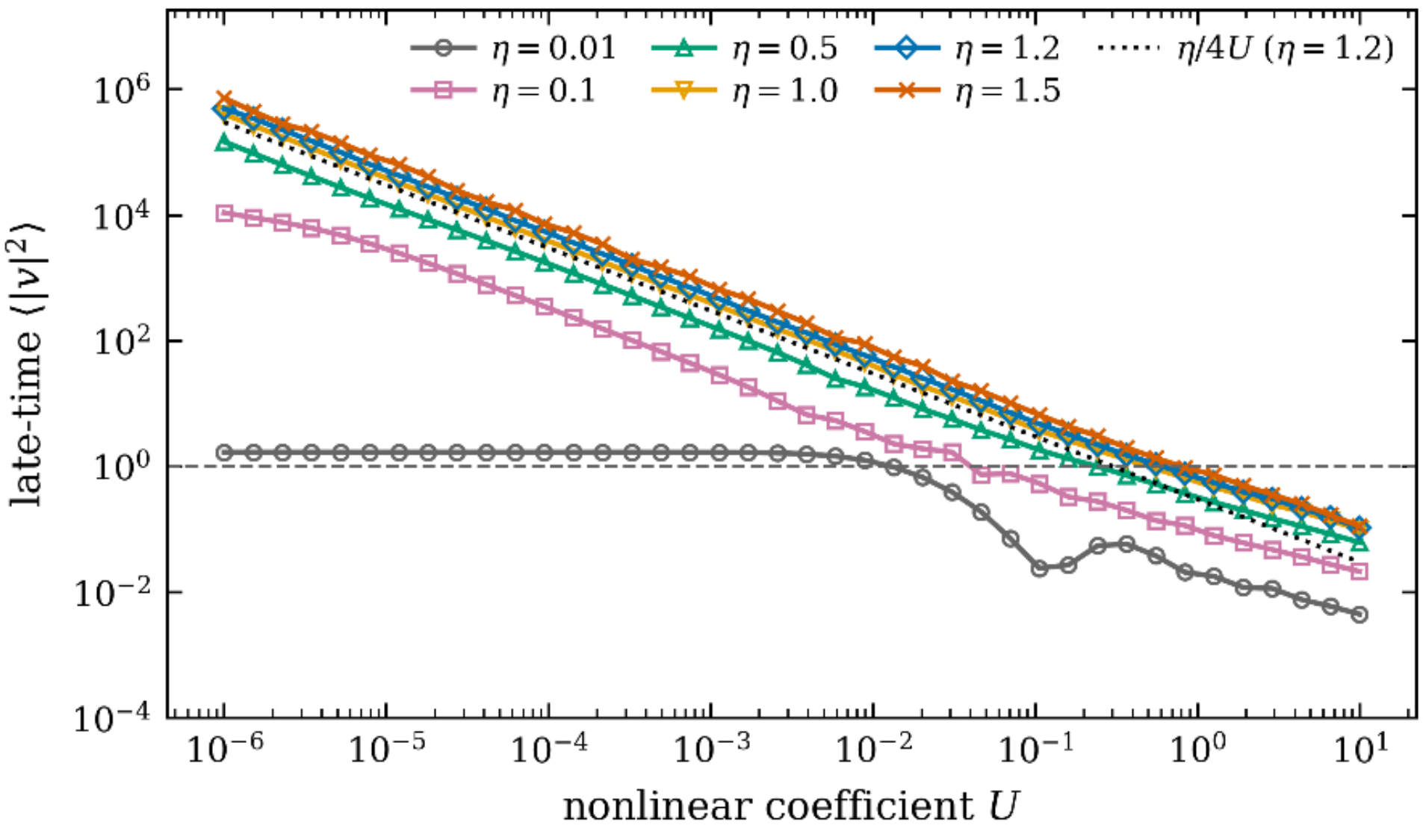

Figure 8. Semiclassical wide-range phase map of non-adiabatic magnon dynamics.

Figure 8 shows the late-time mean occupancy $|v|^2$ as a function of $U$ over a wide range, for several drive intensities $\eta$. At small $U$ the occupancy is large and follows a power law in the drive intensity; as $U$ increases, the trajectories for all $\eta$ converge into a common bounded low-occupancy regime. This is the semiclassical counterpart of the monotonic Lindblad confinement of Fig. 4, extended across a much wider parameter range, and it shows that the confinement is not specific to the few parameter points accessible to the full quantum simulation.

Long-term stability additionally requires sufficient dissipation into the substrate. While intrinsic bulk damping leads to rapid phase divergence, an effective coupling of $\alpha \approx 0.01$, achievable via surface scattering in 20-nm YIG films, provides adequate energy evacuation for bounded evolution over the 100-cycle interval simulated here — the same damping level used for the reference parameter set of Fig. 4.

As quantified in Fig. 4, the reference parameter set ($U_{\text{audit}} \approx 0.55$, $\alpha \approx 0.01$) lies well inside the strongly confined nonlinear regime: the full Lindblad simulation gives $\max_t\langle n\rangle = 0.228$, a late-time mean occupancy $\langle n\rangle = 0.117$, a late-time $P_0 + P_1 = 0.968$ and $P(n \geq 10)$ below $10^{-9}$. The confinement improves monotonically with $U$ across the whole scanned range, so this point is representative rather than optimal: neighbouring values of $U_{\text{audit}}$ between 0.5 and 0.8 give two-level populations within about 1% of it. Because the late-time mean occupancy at this point is $\langle n\rangle \approx 0.12$, the confined state is a low-excitation state within the vacuum–single-magnon manifold rather than a state containing one magnon on average.

Extending the compute window further reveals a consistent ($\eta$, $U$, $\alpha$) trade-off. Raising the substrate coupling to $\alpha \approx 0.01$ extends the coherent compute window from 30 to over 100 cycles at the representative point ($\eta$ = 1.0, $U_{audit}$ = 0.55) without compromising the 22–24 aJ switching energy. Pushing to 1000-cycle coherence requires lower drive intensities: for $\eta \leq 0.8$, the system sustains 1000-cycle coherence with only minimal anharmonicity ($U \approx 0.1$) across all tested $\alpha$. A sharp instability appears near $\eta \approx 1.0$ at low damping ($\alpha$ =0.01); enhancing substrate cooling to $\alpha \geq 0.02$ suppresses this and restores a robust 1000-cycle window even at $\eta$ =1.5 with $U \approx 0.2$. Together these results map the physical trade-off between excitation speed, nonlinear stabilization, and thermal dissipation in this system. The cycle counts in this paragraph are obtained from the semiclassical model. Its prediction of 1000-cycle boundedness at low damping and small anharmonicity ($U \approx 0.1$) is confirmed directly by the full Lindblad calculation of Fig. 3, which reaches a bounded plateau over 1000 cycles at $U = 0.1$, $\alpha = 10^{-3}$; the wider-range trade-offs quoted here (varying $\eta$ and $\alpha$) remain semiclassical.

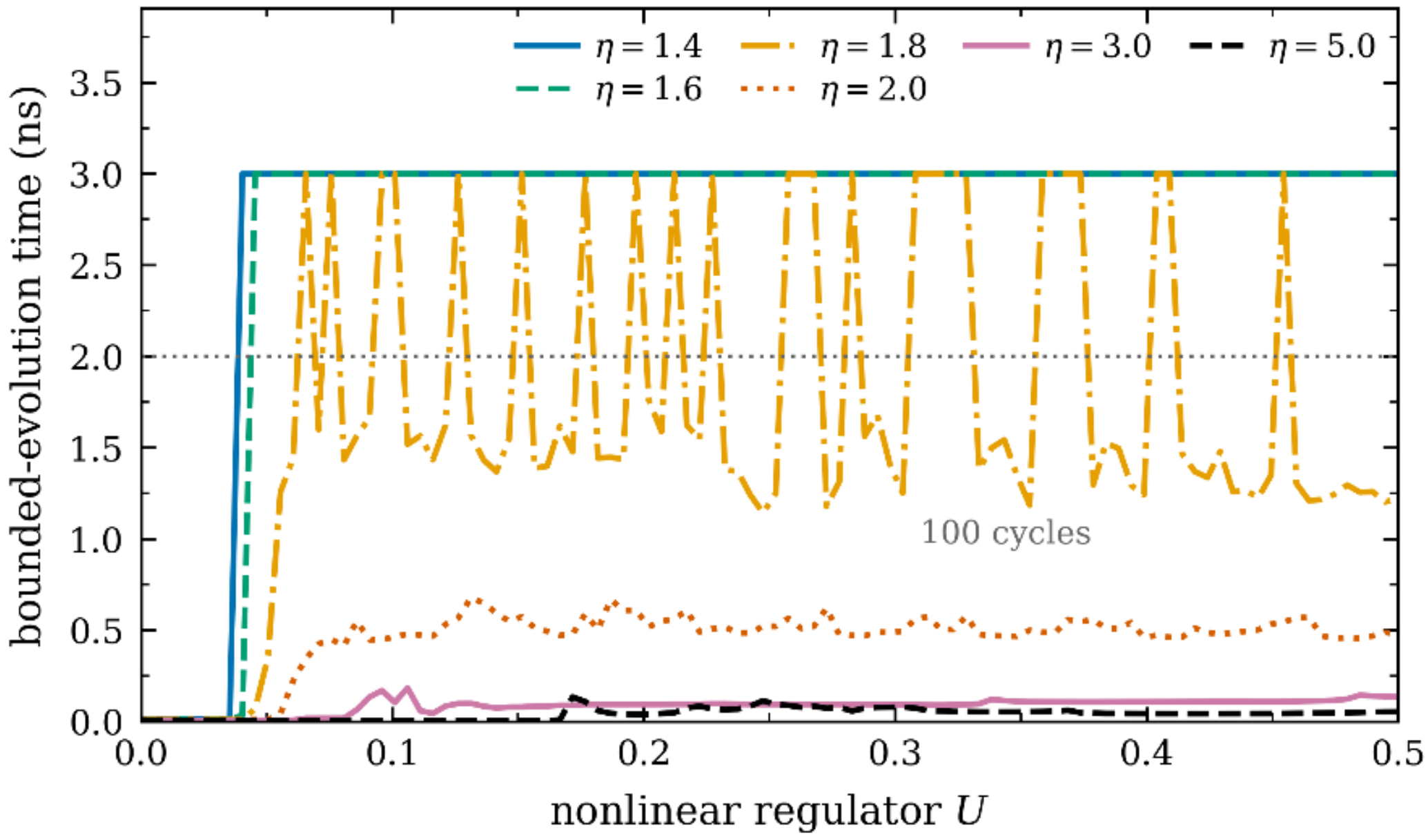

Figure 9. Stability stress-test and Phase Boundaries.

Extending the semiclassical run to 150 cycles (Fig. 9) exposes the phase structure of the $\eta$–$U$ mechanism. A bifurcation appears at the point where the Kerr anharmonicity $U$ (the $\chi^{(3)}$ nonlinearity) starts to offset the parametric drive, and the anharmonicity needed to reach that point, $U_{crit}$, grows with $\eta$. At the reference setpoint $\eta$ = 1.2, with the threshold scaling linearly, every value of $U_{audit}$ above 0.08 gave bounded evolution throughout the test. The map also has a floor in $\eta$. Below $\eta \approx 0.6$ the injected energy fails to exceed the vacuum fluctuations together with the initial nonlinear detuning, the collective mode is never populated, and the run ends in the zero-energy state. Between these limits lies the usable window: the drive has to exceed $\eta \approx 0.8$ for the collective-mode response to be reproducible, while remaining under the divergence limit near $\eta = 1.4$ located in the same test. Setting $\eta = 1.2$ places the system near the centre of that window, trading reliable state transition against sustained coherence.

Between these extremes, at $1.6 \leq \eta \leq 1.8$, the coherence lifetime oscillates in a manner typical of nonlinear resonance: over certain windows the drive phase-locks to the frequency-detuned mode, which is a nonlinear rather than a linear response. Pushing to $\eta = 5.0$ removes the effect altogether; divergence then sets in immediately across the whole $U$-range examined, marking the physical ceiling of the regulation mechanism.

By fixing the operational setpoint at $U_{\text{audit}} = 0.55$, the system operates about four times above the analytically estimated crossover $U_{\text{audit}} \approx 0.12$. This margin protects against drift under manufacturing dispersion; because the Lindblad scan of Fig. 4 shows that the confinement varies smoothly with $U$ throughout this range, moderate deviations from the setpoint degrade the confinement gradually rather than abruptly.

## 6. Thermal and frequency operating regimes

Table 2 presents four operational regimes.  BER for binary pair-phase encoding, evaluated as $1 - \text{erf}[(\pi/2)/(\sqrt{2}\,\sigma)]$, the two-sided probability that the phase error exceeds $\pi/2$. All $\sigma$ values are from full finite-temperature Lindblad runs of the resonantly modulated phase-control protocol at $U_{\text{audit}} = 0.55$, $\eta = 1.2$ and $\alpha = 10^{-3}$, converged in the Fock basis ($N = 80$, 120, 80 and 240 for Regimes I–IV). Note that $\alpha = 10^{-3}$ differs from the value $\alpha = 0.01$ used for the representative operating point of Sections 5.1–5.3; Section 6.3 reports how $\sigma$ depends on the damping. The $P_0{+}P_1$ values for Regimes III and IV are taken from the static effective-pair-drive frequency scan ($U_{\text{audit}} = 0.55$, $\alpha = 10^{-3}$, 20 drive cycles, adaptive Fock basis); the Regime I value is from the 100-level run of Fig. 2.

Table 2. Four operating regimes for a nonlinear magnon mode in YIG:Co

| Regime | $T$ | $\Omega/2\pi$ | $\bar{n}$ | Mechanism | $P_0+P_1$ | Phase $\sigma$ | BER | Cooling |
|---|---|---|---|---|---|---|---|---|
| I Cryogenic Fock | ≲4 K | 50 GHz | ~0 | Fock confinement | 0.95 | 22.4° | $6.1 \times 10^{-5}$ | 4 K closed-cycle |
| II $LN_2$ phase | 77 K | 50 GHz | 31.6 | Pair-phase encoding | — | 29.9° | $2.7 \times 10^{-3}$ | Liquid $N_2$ |
| III RT THz phase | 310 K | 6 THz | 0.65 | Pair-phase encoding | 0.925 | 26.1° | $5.8 \times 10^{-4}$ | None |
| IV RT microwave | 310 K | 50 GHz | 129 | Analog phase mod. | 0.116 | 40.7° | $2.7 \times 10^{-2}$ | None |

### 6.1. Regime II — Liquid-nitrogen phase processor (*T* = 77 K)

At $T = 77$ K and $\Omega/2\pi = 50$ GHz the thermal occupation is $\bar{n} = 31.6$ — a factor of four below the room-temperature value. A full finite-temperature Lindblad calculation at 77 K gives a within-run circular standard deviation of $\sigma = 29.9°$, with a corresponding bit-error rate for binary phase encoding of $2.7 \times 10^{-3}$ ($N = 120$; the population of the highest ten levels stays below $1.1 \times 10^{-7}$). This supersedes the earlier estimate obtained by scaling from the room-temperature result: scaling the within-run circular standard deviation as $\sqrt{\bar{n}}$ from the 310 K value of the same protocol ($\sigma = 40.7°$) gives an estimated $\sigma(77\text{ K}) \approx 40.7° \times \sqrt{(31.6/129)} \approx 20°$, which underestimates the computed 29.9° by about a third because it omits the non-thermal contribution already present at low temperature. The corresponding scaled bit-error rate would be $8 \times 10^{-6}$, far below the computed value. The scaling argument is anchored on the thermally dominated 310 K result and therefore does not include the non-thermal contribution already present at $T \approx 0$ (Section 5.2), where the same diagnostic gives 0.224 rad. Liquid-nitrogen cooling therefore reduces the phase noise substantially relative to room temperature at the same frequency, from 40.7° to 29.9°, but the residual bit-error rate of $2.7 \times 10^{-3}$ is too high for reliable binary encoding without error correction.

### 6.2. Regime III — Room-temperature THz phase processor (*T* = 310 K, 6 THz)

At $\Omega/2\pi = 6$ THz and $T = 310$ K the thermal occupation falls to $\bar{n} = 0.65$, just below the quantum boundary $\bar{n} = 1$ (reached at $\Omega/2\pi \approx 4.5$ THz). A full finite-temperature Lindblad calculation at these conditions (Fig. 10; static effective pair drive at $U_{\text{audit}} = 0.55$, $\alpha = 10^{-3}$, 20 drive cycles, adaptive Fock basis) gives: mean occupancy $\langle n \rangle = 0.245$, pair-correlation amplitude $|\langle \hat{a}\hat{a} \rangle| = 0.393$ and a late-time two-level population $P_0+P_1 = 0.925$, close to the saturation value 0.927 that the scan reaches above 8 THz, while the phase statistics come from a separate calculation, since the static effective pair drive used for the scan is time-independent and therefore has no within-run phase spread; running the resonantly modulated phase-control protocol at the same temperature and frequency gives a within-run pair-phase standard deviation of $\sigma = 26.1°$, larger than the cryogenic 50-GHz value of 22.4° because the residual thermal occupation adds phase noise, and a bit-error rate for binary phase encoding, $1 - \mathrm{erf}[(\pi/2)/(\sqrt{2}\,\sigma)]$, of $5.8 \times 10^{-4}$. The pair-phase transfer coefficient remains −1.000 (kinematic, independent of temperature) and the five-phase control result of Section 5.2 confirms that the pair phase follows the pump phase with a between-phase RMS of $1.3 \times 10^{-3}$ rad at 310 K. The pair correlation $|\langle \hat{a}\hat{a} \rangle|$ varies by 37% across the scan and passes through a minimum near 2 THz (Fig. 10d); what increases monotonically at 6 THz relative to 50 GHz is the pair coherence normalized to the total occupation, because the parametric gain is proportional to $\eta/\alpha$ and the thermal background is suppressed relative to the drive: at $\bar{n} = 0.65$ the normalized pair coherence $|\langle \hat{a}\hat{a} \rangle|/(2\langle n \rangle+1) = 0.264$, against 0.016 at 50 GHz and 310 K — larger by a factor of about sixteen. Regime III is therefore the most favourable room-temperature option: it requires no cryogenic infrastructure, and its phase fluctuations, 26.1°, approach those of the cryogenic Regime I, 22.4°, because the THz drive frequency itself suppresses the thermal background. The drive geometry — a sub-50-fs optical pulse — is exactly that demonstrated by Stupakiewicz et al. [18] for the 22–24 aJ switching event, so the required excitation is experimentally available. Whether a sufficiently isolated 6 THz magnon mode exists to realize the calculated pair coherence is a separate question, and it is not established here: the multimode analysis of Section 6.4 argues against it.

### 6.3. Regime IV — Room-temperature microwave limit ($T$ = 310 K, 50 GHz)

A systematic parameter scan at 50 GHz and 310 K — varying the normalized anharmonicity $U$ from 0.1 to 2.0 and the Gilbert damping $\alpha$ from 0.001 to 0.01, five combinations in all, each integrated for 100 drive cycles with full thermal Lindblad operators — gives within-run pair-phase standard deviations that vary by only a few degrees and show no systematic trend. This is the central negative result of this section: over the combinations tested, $\sigma \approx 40.7°$ is set by $\bar{n} = 129$ at 50 GHz and 310 K rather than by material or drive parameters. This coarse scan is a separate calculation from the converged four-regime phase-statistics campaign reported in Table 2, which was run at $\alpha = 10^{-3}$ only; the $\alpha = 0.01$ campaign was not completed for Regime IV, so the insensitivity established here rests on the coarse scan alone. Reducing $\alpha$ by a factor of ten (from 0.01 to 0.001) leaves $\sigma$ unchanged within 1°; tripling $\eta$ (from 1.2 to 3.0) changes $\sigma$ by less than 5°. The physical reason is that the phase-diffusion rate $D_\varphi = \kappa(2\bar{n}+1)/4|\langle \hat{a}\hat{a} \rangle|^2$ depends on the product $\kappa \cdot \bar{n}$; reducing $\kappa$ (smaller $\alpha$) also reduces $|\langle \hat{a}\hat{a} \rangle|$ so that the ratio barely changes, while $\bar{n}$ is fixed by frequency and temperature alone, independent of both $U$ and $\alpha$. The absolute pair coherence $|\langle \hat{a}\hat{a} \rangle| \approx 0.52$ at this frequency is itself temperature-insensitive and set by the drive rather than by the bath, so the thermal bath leaves the parametric phase-locking mechanism intact and limits only the precision with which the phase can be read out. This insensitivity refers to the normalized model explored here and should not be read as a universal property of magnonic media: it follows from the phase diffusion being set by the thermal occupation, which at 50 GHz and 310 K is already saturated at $\bar{n} = 129$ and cannot be lowered by material or drive optimization at fixed frequency. Regime IV therefore supports analog pair-phase modulation (BER $\approx 2.7 \times 10^{-2}$) but not reliable digital binary encoding, and this limitation cannot be removed by optimizing the material within the Co:YIG family at 50 GHz.

**Dependence on the Gilbert damping.** Because the four regimes above were all computed at $\alpha = 10^{-3}$, a second campaign was run at $\alpha = 0.01$ to separate the temperature dependence from the damping dependence. Stronger damping lowers the phase noise throughout: at 4 K and 50 GHz $\sigma$ falls from 22.4° to 17.9°, and at 310 K and 6 THz from 26.1° to 15.5°, with the corresponding bit-error rate dropping from $5.8 \times 10^{-4}$ to $6.8 \times 10^{-9}$. The zero-temperature control of Section 5.2 gives 12.9° at the same damping, so at $\alpha = 0.01$ the thermal contribution at 6 THz adds only about 3° to the non-thermal floor. This is the expected behaviour of a damped parametric oscillator: a faster relaxation rate re-locks the pair phase to the pump more strongly than it injects thermal noise, since the injection rate $\kappa\bar{n}$ and the restoring rate $\kappa$ grow together while the pump term does not. The trend therefore does not reverse the ordering of the regimes, which is set by $\bar{n}$, but it does mean that the absolute values quoted here are specific to $\alpha = 10^{-3}$ and that a more strongly damped film would perform better. The $\alpha = 0.01$ campaign was not completed for Regime IV, where the thermal occupation $\bar{n} = 129$ requires a Fock basis of $N \gtrsim 400$ and a run time beyond the resources used here.

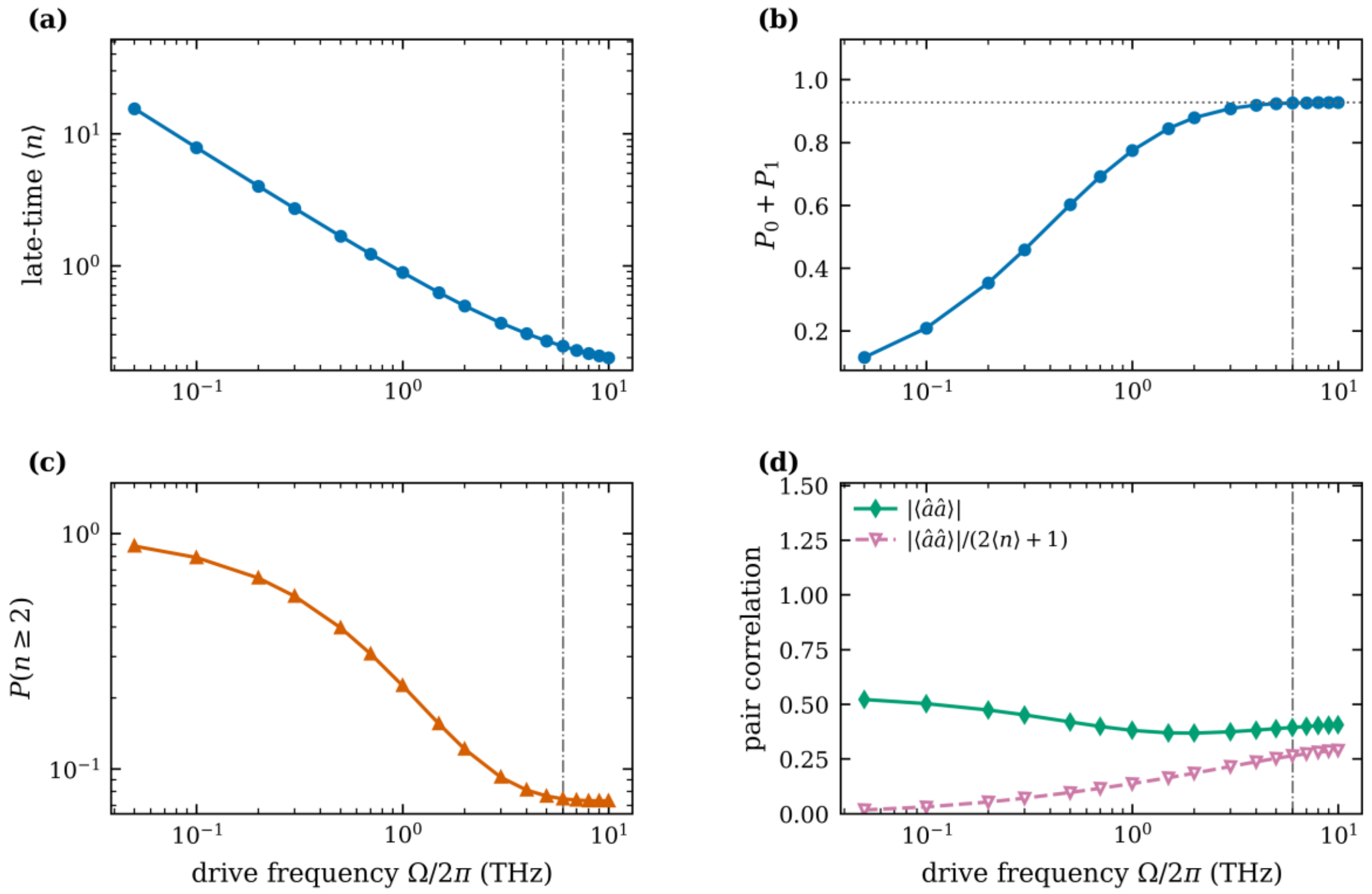

Figure 10. Finite-temperature Lindblad frequency scan ($T = 310$ K, 50 GHz–10 THz; $U_{\text{audit}} = 0.10$ blue, $U_{\text{audit}} = 0.55$ orange). (a) $\langle n\rangle$: indistinguishable for both. (b) $P_0+P_1$: 0.11 at 50 GHz, recovers to $T = 0$ value near 6 THz. (c) Leakage $P(n \geq 2)$ for $U_{\text{audit}} = 0.10$. (d) $|\langle\hat{a}\hat{a}\rangle|$ (blue) between 0.367 and 0.521, with a minimum near 2 THz; normalized coherence $|\langle\hat{a}\hat{a}\rangle|/(2\langle n\rangle+1)$ (orange) grows with frequency.

### 6.4. Damping budget, mode structure and the limits of the single-mode description

With the Kerr term switched off ($U = 0$) nothing constrains the growth, and the occupancy runs away exponentially as exp($t$). The figure of $10^{40}$ quoted in our analysis is a theoretical divergence limit, included to convey how violent the instability of a linear non-adiabatic transition is; a real sample would be destroyed, or the growth would saturate through parasitic channels, long before that. The $\eta$–$U$ mechanism removes the divergence and holds the occupancy in a bounded regime. In the full Lindblad treatment the reference parameter set ($U_{\text{audit}} \approx 0.55$, $\alpha \approx 0.01$) keeps the system at $\max_t\langle n\rangle = 0.228$ with a late-time mean of $\langle n\rangle = 0.117$, a low-excitation state inside the vacuum–single-magnon manifold rather than one holding a magnon on average (Fig. 4).

A comparison with the recent experiments of Karadža et al. [22] is instructive, because the two schemes stabilize different things. Reference [22] achieves a permanent reversal of the magnetization, a bit flip in an open system, through massive excitation of incoherent magnons, which transiently shortens the magnetization vector. The aim here is instead coherent phase control for in-memory processing: the state transition reported above is bounded, confined to a sector dominated by the two lowest Fock levels, and leaves the phase of the $\nu$-mode available to subsequent operations. The two approaches also use nonlinearity in opposite directions. In Ref. [22] nonlinear magnon–magnon scattering is suppressed by compensating the shape anisotropy ($M_{\text{eff}} \rightarrow 0$), so that the injected energy is channelled collectively into the inverted state rather than lost to parasitic modes. Here the Kerr-type nonlinearity is retained and used as a frequency regulator, since without a finite $U$ a drive at $\eta \approx 1$ produces exactly the runaway occupation that destroys confinement. Their stabilization is dissipative, closer to a laser population inversion, with a spin current supplying energy continuously against a damping of $\alpha \approx 0.15$; ours rests on the non-adiabatic resonance windows, and operates at $\alpha \approx 10^{-3}$, where the driven collective-mode state stays spectrally confined and resists rapid thermalization. Where [22] reports incoherent reversal driven by large magnon populations, the $\eta$–$U$ mechanism addresses coherent, phase-controlled evolution inside a confined low-occupancy subspace.

Damping at the level required here has been achieved by doping garnet films: Karadža et al. [22] report $\alpha = 0.00125$ in bismuth-substituted YIG only 10 nm thick, notwithstanding the heavy dopant. That result does not transfer to Co substitution, and the reason is worth stating explicitly. $Bi^{3+}$ is an S-state ion: it enhances the magneto-optical response through the oxygen ligands without opening a new magnon relaxation channel, which is what allows heavy doping at low damping. $Co^{2+}$ supplies the uniaxial anisotropy used here precisely through its unquenched orbital moment and strong spin–orbit coupling, and that same coupling is the dominant relaxation channel responsible for $\alpha \approx 0.15$ in as-grown Co:YIG. Anisotropy and damping therefore share a microscopic origin in this material and cannot be separated by substrate or strain engineering alone; the value $\alpha = 10^{-3}$ used in the calculations below is a target rather than a demonstrated property of Co:YIG, and closing the factor of 150 that separates the two is the principal materials obstacle identified in Section 8. The present model uses Co-doped YIG, with the nonlinearity $U$ supplied by interfacial and strain-induced anisotropy. Two routes towards the assumed damping are proposed. One is structural: growth on GSGG substrates places the film under tensile strain, offsetting the shape anisotropy and reducing nonlinear magnon–magnon scattering. The other exploits the thin-film limit, where at thicknesses of 10 to 20 nm good surface quality and epitaxial integrity have yielded damping an order of magnitude below bulk figures [22]. Both remain routes rather than results, and the damping level assumed in the model is still a target for Co:YIG.

A physical basis for the proposed attojoule operation requires the dissipation channels of a 20-nm Co-doped YIG cell to be separated. We write the effective damping as a sum of three contributions, $\alpha_{eff} = \alpha_{sl} + \alpha_{mm} + \alpha_{def}$. The first is the spin–lattice channel: in insulating garnet the spin system passes energy to the lattice through acoustic phonons.

Cherepanov et al. [23] report an intrinsic linewidth for high-purity YIG at room temperature of roughly 0.15 Oe measured at 35.5 GHz (their Fig. 8), from which the spin–lattice contribution follows as $\alpha_{sl} \approx 1.2 \times 10^{-5}$. Adopting $10^{-3}$ as the target for Co-doped cells therefore leaves room for a hundredfold increase in loss from dopants and surface scattering, a wide engineering margin. Cobalt substitution itself adds local spin–orbit coupling; treating this within the Kasuya–LeCraw model raises $\alpha$ in a controlled way to roughly $5 \times 10^{-4}$.

In macroscopic films, nonlinear scattering into the magnon manifold is a primary source of decoherence. In a cell of this size the exchange gap $\Delta E$ is expected to suppress many-body scattering by detuning the available final states, which would keep $\alpha_{mm}$ small within the operational $\eta$ range [24]. This is a design expectation rather than a computed result: the three-magnon and magnon–phonon channels are not modelled quantitatively anywhere in this work.

Boundary scattering and surface inhomogeneity make up the third contribution. Epitaxial films with roughness under 0.5 nm give $\alpha_{def} \approx 4 \times 10^{-4}$ [25]. Energy relaxation scales inversely with frequency and damping, $\tau_{rel} = (2\pi\alpha f)^{-1}$, so a 50 GHz carrier with $\alpha_{eff} \approx 10^{-3}$ yields $\tau_{rel} \approx 3.2$ ns, or some 160 operation cycles.

The exchange gap of the 20-nm cell ($\Delta E \sim Dk^2$, $k = \pi/L$) places the first non-uniform standing-wave mode 36 GHz above the uniform one, so on the 50 GHz carrier of Regimes I, II and IV the spatial dynamics are dominated by the uniform ($q = 0$) precession mode: no other confined mode falls within the parametric growth rate $h = \eta\Omega/2$ of the carrier, and none inside the damping linewidth, so for that carrier the single-mode approximation is justified. It does not carry over to the 6 THz carrier of Regime III, for the reason set out in the next paragraph. This spatial result is independent of the magnitude of the Kerr nonlinearity $U$ of that mode, which Section 5.1 shows to be $U \approx 6.6 \times 10^{-5}$ for the full uniform mode. The values $U \approx 0.05$–$0.40$ used in the Lindblad calculations are normalized model values; as shown in Section 5.1, they are not realized by any normal mode of a uniformly magnetized cell of this size, so the results below describe the normalized dynamics of a system in which $U$ reaches that range.

**The 6 THz carrier is a different situation.** A uniform mode at 6 THz would require an anisotropy field of 214 T, thirty times the value assumed here, so the 6 THz excitation of Regime III is necessarily a short-wavelength exchange magnon. With the quadratic dispersion $\omega = \gamma(2A/M_s)k^2$ it sits at $k = 2.03 \times 10^9$ m$^{-1}$, a wavelength of 3.1 nm, which is 40% of the way to the zone boundary of the body-centred YIG lattice; at that wavenumber the parabolic approximation itself is 4% high, since $ka = 0.71$ for the 0.35 nm Fe–Fe separation. The consequence for the single-mode description is more serious than either correction. Because the density of confined states grows as $k^2$, the level spacing near 6 THz in this cell is 6.3 GHz on average, while the damping linewidth $\alpha\Omega/2\pi$ is 6.0 GHz: the spacing and the linewidth are equal, the nearest confined mode misses the carrier by 0.25%, and 589 modes lie inside a band of half-width $\eta\Omega/4$ about the carrier; since the parametric growth rate is $h = \eta\Omega/2$, that count is a conservative lower bound on the number of modes above threshold. The target state at 6 THz is therefore one member of a quasi-degenerate manifold rather than an isolated level, and the exchange-gap argument that justifies the single-mode treatment at 50 GHz does not apply to it. Two calculations make the consequence quantitative. First, a linear parametric-stability analysis of the same spectrum: a mode detuned by $\delta$ grows at $\sqrt{(h^2 - \delta^2)} - \kappa/2$ with $h = \eta\Omega/2$, so every mode within $h$ of the carrier is above threshold — one mode on the 50 GHz carrier, where the 36 GHz exchange gap still exceeds the 30 GHz half-width, though by a margin of 1.2 rather than the 2.4 that the narrower band would suggest, but at least 589 modes at 6 THz. Second, a mean-field multimode integration of the same 20 cycles, with self- and cross-Kerr coupling among all modes of the gain band and each mode seeded equally, gives an inverse participation ratio of 1.0 at 50 GHz, the excitation staying entirely in the driven mode, against 53 at 6 THz, where the nominally driven mode retains 0.4% of the total and the three most populated modes together only 21%. On the 6 THz carrier the drive therefore does not address a single mode at all. Third, a full multimode Lindblad calculation, solved in the excitation-number-restricted basis so that the dimension is $C(M + N_{exc}, N_{exc})$ rather than $N^M$. The single-mode case reproduces the scan of Fig. 10 to five digits, and at $M = 4$ the master equation and 400 quantum trajectories agree to 1%. Adding co-driven neighbours at the true level spacing leaves the two-level population intact — it rises from 0.925 to 0.972 — but suppresses the pair correlation that carries the phase: $|\langle\hat{a}\hat{a}\rangle|$ falls from 0.393 for a single mode to 0.128 at $M = 8$, a factor of three, monotonically for $M \geq 4$ and following $|\langle\hat{a}\hat{a}\rangle| \propto M^{-1.31}$ to within 4%. The mechanism is not leakage: for a spatially uniform pump the overlap integral vanishes unless the two mode functions coincide, so the drive excites each mode within itself and the modes are coupled only through the cross-Kerr term, which detunes the target mode out of parametric resonance. Regime III is nevertheless reported here on the same single-mode footing as the other three, so that the four are computed identically and the comparison between them is meaningful; for that carrier this is an assumption, and the calculations above indicate that a real film would show a substantially weaker pair coherence than the single-mode numbers suggest. Extrapolating the fitted power law to the 589 modes of the gain band would give $|\langle \hat{a}\hat{a} \rangle|$ of order $10^{-3}$, but that is an extrapolation two decades beyond the computed range and is quoted as an indication of direction, not as a result. What is unaffected is the position of Regime III on the $\hbar\Omega/k_B T$ axis, since the thermal occupation $\bar{n} = 0.65$ follows from the frequency and the temperature alone.

Energy leaving the confined magnon state reaches the lattice through spin–lattice relaxation, at a rate set by $\alpha$, and appears there as acoustic phonons. The GGG substrate serves as the principal heat sink and carries them away.

The semiclassical stress test of Fig. 9 bounds the operating point from below rather than from parameter dispersion: for the static-drive protocol, bounded evolution is observed for $U_{\text{audit}} > 0.08$ at $\eta = 1.2$. No dispersion, yield or manufacturing-success-rate figure is computed or claimed here.

Spectral separation is what makes the $\eta$–$U$ mechanism robust. Micromagnetic analysis (MuMax3-equivalent) of a 20×20×10 nm$^3$ Co-doped garnet cell puts the exchange gap $\Delta E$ in the range $10^{-23}$–$10^{-22}$ J, specifically $\Delta E \sim 6.9\times10^{-23}$ J, in line with the exchange-mode quantization energy $\Delta E \sim Dk^2$ with $D$ the exchange stiffness and $k \sim \pi/L$ set by the confinement length. That is roughly 1.7 times the largest energy injected by the non-adiabatic pulse. The figure is an order-of-magnitude estimate of how far the localized computational mode sits from higher-order spin-wave excitations and should be read as such.

Micromagnetic simulations give a stable uniform precession at $\eta \leq 0.8$; at the Hamiltonian optimum $\eta = 1.2$ the simulations suggest caution regarding parasitic spatial spin-wave excitation, which the purely temporal single-mode model does not capture. The 1000-cycle stability result of Fig. 3 is established by the full Lindblad calculation (Section 5.1) for the temporal single-mode model and does not depend on the micromagnetic exchange-gap estimate. The two analyses address complementary aspects: the Lindblad calculation establishes long-time temporal boundedness; the micromagnetic analysis establishes spatial single-mode validity.

## 7. Discussion

The dimensionless ratio $\xi = \eta/U$ measures whether a non-adiabatic transition remains dynamically stable under strong excitation, and it is used here as a bookkeeping variable rather than as a universal threshold. Analogous crossover behaviour in independently studied time-modulated and nonlinear resonant systems, including coupled nonlinear metasurfaces and time-modulated resistive metasurfaces [26, 27], is analysed within the same framework in companion work [4].

The frequency scan makes the role of the control ratio explicit. Table 3 lists the multi-THz end of the scan at 310 K. Across 6–10 THz the absolute pair correlation $|\langle \hat{a}\hat{a} \rangle|$ changes by about 3%, confirming that it is set by the drive rather than by the bath, while the late-time two-level population is already 0.925 at 6 THz and increases only marginally thereafter, saturating at 0.927. The approach to confinement is nevertheless gradual rather than abrupt: the same scan gives 0.908 at 3 THz, 0.919 at 4 THz and 0.923 at 5 THz, so any stated onset frequency reflects a chosen threshold on $\hbar\Omega/k_B T$, and the residual improvement from 6 to 10 THz is small compared with the change between 50 GHz and 6 THz. This is the practical content of organizing the four regimes by $\hbar\Omega/k_B T$ — at room temperature, moving from 50 GHz to 6 THz changes the operating regime, whereas pushing further into the THz band buys little.

The absolute error rates deserve a comment. None of the four regimes reaches uncorrected error-free operation: the raw bit-error rates run from $6 \times 10^{-5}$ at 4 K to $3 \times 10^{-2}$ at room temperature and 50 GHz. That does not by itself disqualify the scheme. Coherent optical links operate routinely at raw pre-decoding error rates of this order and recover error-free data with forward error correction: hard-decision codes with 7% overhead have thresholds near $4 \times 10^{-3}$, and soft-decision codes with 20% overhead near $2 \times 10^{-2}$ [28]. On that scale Regimes I and III sit comfortably inside the correctable range, Regime II near the hard-decision threshold, and Regime IV at the edge of what even soft-decision coding can recover. These thresholds come from a different technology and are quoted for orientation only, not as a specification for magnonic logic; the comparison is made because a raw error rate of $10^{-4}$–$10^{-3}$ is a coding problem rather than a physical obstruction, whereas the $3 \times 10^{-2}$ of Regime IV is close to being one.

Table 3. Multi-THz end of the finite-temperature Lindblad frequency scan of Fig. 10 (static effective pair drive, $T = 310$ K, $U_{audit} = 0.55$, $\alpha = 10^{-3}$, 20 drive cycles, adaptive Fock basis: $N = 45$ above 0.7 THz rising to 209 at 50 GHz, each point repeated at $1.2N$). $\bar{n}$ is the thermal occupation of the undriven mode, $P_0+P_1$ the late-time two-level population, $\langle n \rangle$ the late-time mean occupation and $|\langle \hat{a}\hat{a} \rangle|$ the absolute anomalous pair correlation.

| $\Omega/2\pi$ | $\bar{n}$ | $P_0+P_1$ | $\langle n \rangle$ | $\lvert\langle \hat{a}\hat{a} \rangle\rvert$ |
|---|---|---|---|---|
| 6 THz | 0.65 | 0.925 | 0.245 | 0.393 |
| 7 THz | 0.51 | 0.926 | 0.229 | 0.398 |
| 8 THz | 0.41 | 0.927 | 0.217 | 0.401 |
| 9 THz | 0.33 | 0.927 | 0.208 | 0.404 |
| 10 THz | 0.27 | 0.927 | 0.201 | 0.406 |

The two-level population is above 0.92 throughout and $|\langle \hat{a}\hat{a} \rangle|$ varies by 3% over this range, so the confinement gain from raising the drive frequency above 6 THz is marginal (see Table 3).

$\eta$ takes different values in the earlier and later parts of this work, and the shift is one of description rather than of physics: the treatment moves from a purely Hamiltonian account to a micromagnetic one. In the Hamiltonian part the switching energy of 22–24 aJ emerges as a dynamical limit for an idealized, dissipationless vacuum, and $\eta = 1.2$ is the optimum there for near-instantaneous state formation.

The bosonic Hamiltonian of Eq. 4 connects to the micromagnetic Landau–Lifshitz–Gilbert description through the Holstein–Primakoff transformation. For small precession angles the equations of motion of the $\eta$–$U$ mechanism take the form of an LLG equation carrying a time-dependent parametric drive and a nonlinear saturation term. What this fixes is the formal correspondence between the reduced bosonic description and small-angle LLG dynamics; it says nothing about whether the required range of the nonlinear parameter is attainable in Co:YIG.

Explicitly, the magnetization operators are written in bosonic form as: $S^+ = \sqrt{2S - \hat{a}^\dagger\hat{a}}\,\hat{a}$, $S^z = S - \hat{a}^\dagger\hat{a}$. Provided the mode occupation obeys ($\langle \hat{a}^\dagger \hat{a} \rangle \ll 2S$), the Heisenberg equation $\dot{\hat{a}} = i\,[\hat{H}, \hat{a}]$ reduces to the linearized LLG equation: $\partial m/\partial t = -\gamma m \times H_{eff} + \alpha m \times \partial m/\partial t$ . Here the drive $\eta$ enters as a time-dependent contribution to $H_{eff}$, and $m = M/M_s$ is the magnetization normalized to its saturation value $M_s$. The gyromagnetic ratio $\gamma$ fixes the precession rate of the moment in an applied field, while $H_{eff}$ gathers the bias, anisotropy and exchange contributions along with the time-dependent drive $\eta(t)$ responsible for the non-adiabatic transition. Gilbert damping appears in the final term through the dimensionless constant $\alpha$ and carries the energy given up to the lattice. The reduced bosonic description therefore tracks the actual rotation of the magnetization vector, set by the competition between parametric torque and nonlinear relaxation.

The switching-energy figure of 22–24 aJ follows from the reported absorbed-energy density and the cell volume (Section 3), not from the magnon energy quantum at 50 GHz. Small numerical fluctuations ($< 0.2\%$) observed during the convergence verification do not affect the boundedness of the confined state over the simulated interval.

The later sections use a micromagnetic-equivalent treatment built on Bogoliubov–de Gennes dynamics, which admits stochastic effects, Gilbert damping $\alpha$ and thermal exchange with the substrate. The Hamiltonian burst at $\eta = 1.2$ remains appropriate for single-cycle operation, while sustained coherence over 1000 cycles in a dissipative medium is reached at $\eta \leq 0.8$. Boundedness over 1000 drive cycles is verified directly by the full Lindblad calculation of Fig. 3 at $U = 0.1$ and $\alpha = 10^{-3}$, whereas the $\eta$-dependence quoted here comes from the semiclassical model. On this reading the two levels of description are consistent: the Hamiltonian sets single-cycle behaviour, and the semiclassical treatment governs long-term stability through dynamical localization.

Any device-level comparison should be read with caution, since the aJ figure refers to the absorbed energy of an isolated switching event and not to a system-level energy-per-operation, which would additionally include drive generation, readout, interconnect and reset. With that caveat, the isolated-event switching energy is well below the per-operation energy of charge-based CMOS logic, in which interconnect and memory access dominate the budget [8].

As established in Section 5.1 (Fig. 4), keeping the occupancy bounded and confining it deep inside the low-excitation manifold are distinct requirements: the confinement tightens monotonically with $U$, and the representative combination ($U_{\text{audit}} \approx 0.55$, $\alpha \approx 0.01$) sits well inside that regime, with about 97% of the late-time population in the two lowest Fock states.

Requiring strong Kerr-type nonlinearity, that is a large normalized regulator $U$, together with damping as low as $\alpha \approx 10^{-3}$ sets a hard materials problem, and we see two complementary routes. The first is interfacial: concentrating the cobalt at the interface rather than through the bulk supplies surface-induced anisotropy while leaving the bulk magnon mode in low-loss YIG. The second is strain: epitaxial YIG:Co grown on GSGG or substituted GGG under controlled tensile strain acquires anisotropy from the strain itself, which permits a lower dopant concentration and with it a damping constant nearer the target, along the lines of the benchmarks reported by Karadža et al. [22].

An experimental test of the $\eta$–$U$ mechanism could proceed in two stages. The first uses time-resolved magneto-optical Kerr effect or micro-focused Brillouin light scattering to follow phase persistence and relaxation in the non-adiabatic regime, with a power sweep to locate the 22–24 aJ excitation threshold and to establish where nonlinear stabilization by $U$ sets in. The second applies spin-wave spectroscopy to 20-nm cells, confirming the single-mode spectrum and the exchange gap $\Delta E$ estimated above.

Relative to established magnonic computing — spin-wave logic and magnonic crystals [5–7], which rely on propagation and interference over micrometre distances and are thereby limited in clock frequency and footprint — the mechanism studied here is a localized, single-mode nonlinear transformation rather than a wave-interference scheme. The same distinction separates it from all-optical ultrafast switching [29], where the central difficulty is heat production: the confinement mechanism suppresses multi-mode scattering during the quench, while the 22–24 aJ figure is taken from the experimental absorbed-energy density and is not derived from the confinement model. Whether these single-cell properties can be organized into a large, individually addressable array — which would additionally require readout, inter-cell coupling and fan-out not treated here — is a device-engineering question beyond the scope of the present single-mode model.

## 8. Conclusions

The principal result of this work is a criterion together with a demonstration that the criterion is not enough. The Kerr coefficient of a confined magnon mode is fixed by its net spin content — $U = \omega_a/(2N_{\mathrm{eff}}\Omega)$ with $N_{\mathrm{eff}} = M_s V/(g\mu_B)$, equivalently $K = \hbar\gamma^2 K_u/(M_s^2 V)$ — while the drive is bounded from below by the damping, $\eta > 2\alpha$, and from above by the requirement that the saturated occupation $\eta/(4U)$ stay under unity, $\eta < 4U$. A window $2\alpha < \eta < 4U$ requires $\alpha N_{\mathrm{eff}} < \omega_a/\Omega$, which for a mode driven at its own anisotropy frequency is $N_{\mathrm{eff}} < 1/\alpha$. The lower bound is exact and the upper one is a mean-field estimate, so this is a screening criterion rather than a theorem. Thermal stability $K_u V \geq 40 k_B T$ bounds the mode volume from below and therefore bounds $N_{\mathrm{eff}}$ from below as well. For a uniformly magnetized Co-doped YIG cell the condition is met at both 50 GHz and 200 GHz by low-damping garnet, and the mechanism is still unavailable: the window it opens sits at $\eta \approx 10^{-4}$, and the Kerr coefficient the cell delivers is three orders of magnitude below the range at which confinement appears. A material condition of this kind can be satisfied while the operating point it admits lies outside the physical regime the mechanism requires, and that is the situation here. What survives is the pair-phase transfer, whose coefficient is kinematic; the ordering of its noise floor across the four regimes is set by $\hbar\Omega/k_B T$, while the absolute magnitude also depends on the damping and the drive protocol. The four regimes below map that noise floor, computed by full Lindblad calculations with explicit thermal operators.

***Regime I ($T \lesssim 4$ K, 50 GHz).*** The Kerr anharmonicity confines more than 95% of the late-time population to the two lowest Fock states over 1000 drive cycles. The pair-phase standard deviation is 22.4° and the switching energy is 22–24 aJ per cell. Requires a 4 K closed-cycle cryostat, not a dilution refrigerator.

***Regime II ($T = 77$ K, 50 GHz).*** Liquid-nitrogen cooling reduces $\bar{n}$ to 31.6 and the pair-phase standard deviation to 29.9°, down from 40.7° at the same frequency at room temperature, with a bit-error rate of $2.7 \times 10^{-3}$, which requires error correction for digital use. Most accessible intermediate target: no helium, standard 50-GHz microwave technology.

***Regime III ($T = 310$ K, 6 THz).*** When $\hbar\Omega \gtrsim k_B T$ ($\Omega/2\pi \gtrsim 4.5$ THz at 310 K), $\bar{n}$ falls to 0.65, $|\langle \hat{a}\hat{a} \rangle|$ reaches 0.393, the late-time two-level population recovers to $P_0+P_1 = 0.925$, and the pair-phase standard deviation is 26.1°, giving a raw bit-error rate of $5.8 \times 10^{-4}$ — the lowest of the room-temperature options, but still requiring error correction for digital use. This is the room-temperature phase-encoding regime; it corresponds to the sub-50-fs optical pulse geometry of Stupakiewicz et al. [18] and requires no cryogenic infrastructure.

***Regime IV ($T = 310$ K, 50 GHz).*** The pair-phase standard deviation reaches 40.7°, the largest of the four regimes, and across the five combinations tested it changes by only a few degrees. The limit is set by the thermal occupation $\bar{n} = 129$ and, within that range, is not lifted by material or drive optimization at fixed frequency. Supports analog phase modulation (BER $\approx 2.7 \times 10^{-2}$) but not reliable digital binary encoding.

***Material recommendation.*** For quantum memory compatible with existing 4 K platforms: Regime I, requiring $\alpha \approx 10^{-3}$, which is a target for optimized Co:YIG films rather than a demonstrated value. For room-temperature magnonic signal processing with femtosecond laser sources: Regime III — of the material quality assumed throughout Section 6, the anisotropy $K_u \approx$ $5{\times}10^5$ J m$^{-3}$ lies within the range reported for Co-substituted garnet films, while $\alpha = 10^{-3}$ does not and remains a target, and the pulse geometry is established. The nonlinear coefficient used in the Regime III calculations ($U_{\text{audit}} = 0.10$–$0.55$, that is $U = 0.05$–$0.275$) is, however, three orders of magnitude larger than the value obtained for a uniformly excited 20 × 20 × 10 nm³ cell with these material parameters ($U \approx 6.6 \times 10^{-5}$, Section 5.1), and therefore presupposes a mode far smaller than any that exists in the cell: a normal-mode analysis of the same geometry finds no localised mode, the least extended one still spanning about a quarter of the cell (Section 5.1). The Regime III calculations are accordingly normalized-model results rather than a prediction for a uniformly magnetized Co:YIG cell of this size. The composition and geometry of commercially available films are consequently a promising starting point rather than a sufficient condition, and experimental confirmation of the required nonlinearity at the operating frequency remains necessary. For applications where liquid nitrogen is acceptable: Regime II offers digital phase encoding at 50 GHz without liquid helium. Regime IV remains viable for analog phase modulation in room-temperature microwave environments but is not recommended for digital applications.

**Where the specification fails.** Selecting an operating regime fixes the frequency and the temperature; it does not fix the geometry, the material parameters or the drive schedule, which are specified here. Table 4 collects a specification for the most conservative case, Regime I on a 50 GHz carrier, using only numbers that are either measured or follow from the material parameters. The film thickness and the lateral cell size are at the limits of routine liquid-phase epitaxy and electron-beam patterning rather than at any physical limit; the anisotropy is set to place the uniform mode at the carrier frequency, which for $M_s = 1.4 \times 10^5$ A m$^{-1}$ means $K_u = 1.25 \times 10^5$ J m$^{-3}$ rather than the $5 \times 10^5$ J m$^{-3}$ of the switching experiment; and the damping is the optimized value, two orders below the as-grown Co-substituted films.

Table 4. Where the specification of a Regime I cell on a 50 GHz carrier fails. Entries in the third column marked as computed follow from the listed material parameters. The table is assembled not as a design target but as a refutation: the last row records the quantity the material does not supply, and it falls three orders of magnitude short of the range the calculations require.

| Quantity | Value | Origin |
|---|---|---|
| film thickness | 10 nm | LPE / PLD, routine |
| lateral cell | 20 × 20 nm | e-beam patterning limit |
| cell volume | 4000 nm³ | geometry |
| saturation magnetization | $1.4 \times 10^5$ A m$^{-1}$ | YIG at 300 K |
| $Fe^{3+}$ ions per cell | $8.4 \times 10^4$ | 40 per 1.90 nm³ unit cell |
| spin content $N_{eff} = M_S V/(g\mu_B)$ | $3.0 \times 10^4$ | net moment / $g\mu_B$ |
| uniaxial anisotropy | $1.25 \times 10^5$ J m$^{-3}$ | set to place the mode at 50 GHz |
| carrier $\Omega/2\pi$ | 50 GHz | mode frequency |
| parametric pump $2\Omega/2\pi$ | 100 GHz | microwave |
| Gilbert damping α | $10^{-3}$ | target for an optimized film; 0.15 as-grown Co:YIG |
| thermal stability $K_u V/k_B T$ | 117 at 310 K | computed |
| switching energy | 24 aJ | 6 J cm$^{-3}$ × V, Ref. [18] |
| ramp time $1/(\eta\Omega)$ | 2.7 ps | $\eta = 1.2$ |
| clock at 100 cycles | 0.5 GHz | $\Omega/2\pi N_{cyc}$ |
| Kerr coefficient U | $6.6 \times 10^{-5}$ | $\omega_a/(2 N_{eff}\,\Omega)$ — three orders below the model range |

One entry of that specification is not met, and is reported as such rather than as a design target. The Kerr coefficient that the cell delivers, $6.6 \times 10^{-5}$, is three orders of magnitude below the range the calculations require, and Section 5.1 shows that no geometry within the magnonic setting recovers it: shrinking the cell raises $U$ as $1/V$ and lowers the switching energy in proportion, but the thermal-stability limit $K_u V \geq 40 k_B T$ stops the trade at a 7 nm cube and $U = 7.7 \times 10^{-4}$. Second, the 6 THz carrier of Regime III cannot be a uniform mode of this material at all: a uniform mode at 6 THz would need an anisotropy field of 214 T, thirty times the value assumed here, so the 6 THz excitation is necessarily a short-wavelength exchange magnon with $k \approx 2.0 \times 10^9$ m$^{-1}$, a wavelength of 3.1 nm. Section 6.4 shows that at that wavenumber the confined level spacing, 6.3 GHz, has fallen to the damping linewidth, so the target state is one of a quasi-degenerate manifold and the single-mode treatment of Regime III is an assumption rather than a derived result. Its thermal occupation, and with it the position of Regime III on the $\hbar\Omega/k_B T$ axis, is unaffected by this; what is affected is the single-mode description, which Section 6.4 quantifies: a full multimode Lindblad calculation with the neighbouring modes included suppresses the pair correlation of the driven mode by a factor of three already at eight modes, and the suppression is still growing there. Both points are stated here rather than deferred because they define what an experiment would have to overcome, and neither changes the ordering of the four regimes, which follows from $\hbar\Omega/k_BT$ alone.

A materials criterion. The specification above fails on one entry, and the model itself says which one and by how much. Two inequalities have to hold at once. The parametric growth rate must exceed the amplitude decay, $h > \kappa/2$ with $h = \eta\Omega/2$ and $\kappa/2 = \alpha\Omega$, that is $\eta > 2\alpha$; and the saturated occupation $\eta/(4U)$ must stay below one, that is $\eta < 4U$. The first of these is exact, being the linear parametric threshold. The second is not: $\eta/(4U)$ is the mean-field balance of Section 5.3, and its accuracy is protocol-dependent. At the resonantly modulated operating point it predicts 1.09 against the Lindblad value 1.213, an agreement to 11%; under the static effective drive it overestimates badly, the 1000-cycle calculation of Fig. 3 giving $\langle n\rangle = 0.176$ and $P_0+P_1 = 0.952$ where the balance predicts 3. The window $2\alpha < \eta < 4U$ is therefore a screening construction at mean-field level rather than a proved necessary condition, and what follows should be read as a rule for ranking candidate media, not as a theorem excluding them. On that footing a window requires $\alpha < 2U$. Writing $U = \omega_a/(2N_{eff}\Omega)$ turns this into a statement about how much magnetic material the mode may contain, $\alpha N_{eff} < \omega_a/\Omega$, which for a mode driven at its own anisotropy frequency ($\omega_a = \Omega$) reduces to $N_{eff} < 1/\alpha$. The material estimates below are made at $\omega_a = \Omega$, the anisotropy being chosen in each case to place the mode at the stated carrier, so the reduced form applies to them; the general form is the one to use for any medium in which the drive is detuned from the anisotropy frequency, as it is for the illustrative cell of Section 5.1, where $\omega_a/\Omega = 4$. That form is the one used here, because it is independent of the normalization chosen for the Kerr term and refers only to two measurable quantities, the Gilbert damping and the spin content of the mode. Thermal stability pushes in the opposite direction, $K_uV \geq 40k_BT$, and the two together close the design space. For a mode at 50 GHz, which needs $K_u = 1.25 \times 10^5$ J m$^{-3}$, stability requires $V \geq 1370$ nm$^3$, hence $N_{eff} = 1.0 \times 10^4$ and $\alpha < 9.7 \times 10^{-5}$. Raising the anisotropy to $5 \times 10^5$ J m$^{-3}$, which places the mode at 200 GHz, shrinks the stable volume to 342 nm$^3$ and relaxes the requirement to $\alpha < 3.9 \times 10^{-4}$. Both are met by low-damping garnet: the best bulk YIG single crystals reach $\alpha \approx 4 \times 10^{-5}$, and high-quality films are within a factor of two of the 50 GHz bound. The criterion is therefore satisfied, and it is precisely here that its not being sufficient decides the outcome. At $\alpha = 3.9 \times 10^{-5}$ the admissible drive interval is $7.8 \times 10^{-5} < \eta < 1.9 \times 10^{-4}$ at 50 GHz and $7.8 \times 10^{-5} < \eta < 7.7 \times 10^{-4}$ at 200 GHz. Neither is empty, and neither is useful: both lie three to four orders of magnitude below the $\eta = O(1)$ that defines the non-adiabatic drive on which the mechanism is built. A medium can pass the material test and still admit only a drive so slow that the transition it is supposed to stabilize does not occur. The one material combination that fails the test outright is the practically relevant one: as-grown Co:YIG at $\alpha = 0.15$ would require $N_{eff} < 7$, a mode occupying less than 1 $nm^3$, which is the same contradiction seen from the other side — the Co substitution that supplies the anisotropy is what raises the damping. The criterion is stated in the form $\alpha N_{eff} < \omega_a/\Omega$ because it allows any other candidate medium to be tested directly against the scheme, whether or not that medium is a garnet. It applies to the confined regimes only: as Section 5.2 shows, the pair-phase transfer is kinematic and holds unchanged down to $U = 0.01$, so phase readout requires nothing beyond a parametric drive that exceeds the damping, $\eta > 2\alpha$. Of the two encoding mechanisms examined here, the pair-phase observable is the one a garnet cell can support in the model.

**Experimental strategy**. The cobalt content is the most direct compositional parameter, but the criterion indicates that it is not the appropriate one to vary. Written out, $U = \omega_a/(2N_{eff}\Omega)$; a mode driven at its own frequency has $\Omega = \omega_a$ and therefore $U = 1/(2N_{eff})$, independent of the anisotropy altogether. Cobalt fixes where the mode sits in frequency, not how nonlinear it is, while it contributes to the damping directly. The empirical trend is in the same direction: going from pure YIG to the doped material raises $K_u$ by about three orders of magnitude but $\alpha$ by nearly four, so the ratio that matters degrades by roughly a factor of six. Composition should therefore be reduced to whatever is needed to place the mode, and not increased beyond that.

Three variables are worth scanning instead. The first is the mode volume, which enters through $N_{eff} < 1/\alpha$ and is bounded from below by thermal stability, so that the two together define a window rather than a direction. The second is the source of the anisotropy: interfacial anisotropy from a capping layer, or strain from a lattice-mismatched substrate, sets $K_u$ without introducing a fast-relaxing ion into the garnet, and is the only route we can identify that separates the two properties shown above to share an origin. The third is temperature. The $Co^{2+}$ contribution to $\alpha$ is a slow-relaxation process peaked where the ionic relaxation rate matches the mode frequency, so moving the operating point away from that peak lowers the damping at fixed composition and leaves $K_u$ essentially untouched; it is the least demanding of the three tests experimentally.

The decisive measurement is a determination of the linewidth. Extracting $\alpha$ from ferromagnetic resonance and estimating the mode volume from the film thickness and lateral patterning gives $N_{eff} = M_s V/(g\mu_B)$, and the sample either satisfies $N_{eff} < 1/\alpha$ or it does not. At the net spin density of 7.5 $nm^{-3}$ that follows from $M_s = 1.4 \times 10^5$ A $m^{-1}$, as-grown Co:YIG at $\alpha = 0.15$ admits $N_{eff} <$ 7, a mode of under 1 $nm^3$, which no patterned structure will meet; at $\alpha = 10^{-3}$ the bound is $N_{eff} < 10^3$, a volume of 132 $nm^3$ or a 5.1 nm cube; at the best reported garnet damping, $\alpha = 3.9 \times 10^{-5}$, it reaches 3400 $nm^3$, a 15 nm cube. Taken together with the thermal-stability floor of 1370 $nm^3$ at 50 GHz, the last of these leaves an open interval of mode volumes rather than a closed one, and that interval is comfortably inside what electron-beam patterning reaches. That progression, rather than any composition series, is the sequence against which a candidate structure should be assessed.

What does not depend on the material. Two results of this work do not depend on the material. The first is the control ratio: $\hbar\Omega/k_B T$ determines which information-encoding mechanism remains physically accessible in a driven magnon mode, and the medium enters only through how far along that ratio a given film and drive can be pushed. The second is the mean-field screening condition that follows from the drive window $2\alpha < \eta < 4U$: a window requires $\alpha < 2U$, and since $U$ is fixed by the spin content of the mode this reads $N_{eff} < 1/\alpha$. It screens rather than excludes, and it is not sufficient in any case — the window can be non-empty and still lie far below the non-adiabatic drive the mechanism requires, as it does at 200 GHz. Any candidate medium, garnet or otherwise, can be tested against the scheme by comparing its damping with the inverse spin content of the mode it supports. For the garnet cell examined here that comparison is passed at both carriers, and the mechanism is nevertheless unavailable: the drive window the criterion admits lies near $\eta \approx 10^{-4}$, three to four orders of magnitude below the $\eta = O(1)$ that makes the transition non-adiabatic, and the Kerr coefficient the cell delivers is three orders below the range at which confinement appears. What future experimental work would establish is therefore not whether the inequality can be met — it can — but whether any medium meets it at a spin content small enough for the admissible drive to remain non-adiabatic.

## 9. Supplementary Materials

The Supplementary Material is organized into eight sections. Supplement 1 gives the numerical models, the open-system and Hamiltonian conventions of each protocol, the observables, and the numerical provenance of every calculation. Supplement 2 gives the relation between the Lindblad and Bogoliubov descriptions: the exact linear structure and the required input-noise term, the mean-field nonlinear closure, and the damped norm diagnostic $|u|^2 - |v|^2 = \exp(-2\alpha\Omega t)$ (Figure S1). Supplement 3 reports the drive-protocol audit isolating the effect of the pair-drive implementation at matched conventions (Figure S2). Supplement 4 gives the matched Bogoliubov–Lindblad bridge, the parity structure of the pair drive, and the anomalous-correlator convention audit. Supplement 5 reports the pair-phase control tests: the stationary five-phase transfer and the dynamical phase-step response (Figure S3). Supplement 6 states the origin of the 22–24 aJ energy scale. Supplement 7 is a reproducibility checklist. Supplement 8 documents the finite-temperature frequency scan at 310 K from which Fig. 10, Table 3 and the Regime III and IV entries of Table 2 are taken, including the basis-convergence statement for the low-frequency end of the scan (Figure S4). The accompanying software archive contains the protocol-audit, five-phase and step-response scripts, the publication-figure generator, the exploratory notebooks, and the CSV summaries and figures.

**Competing Interests.** The author declares no competing interests.

**Data Availability.** The supplementary materials, the numerical data, figure-generation scripts and checklist are available at https://github.com/alexandermtishin-art/Nonlinear-magnon-dynamics-in-Co-doped-YIG-four-operating-regimes- . The archive contains all Python/QuTiP scripts for Figs. 1–10 and Tables 1–4, CSV outputs, and a SHA256 manifest.

## Declaration of Generative AI in Scientific Writing

During the preparation of this work, the author used ChatGPT, Claude and Google Gemini for language editing, notation and consistency checks, critical-review assistance, and the preparation and verification of auxiliary Python scripts and figures. All scientific hypotheses, interpretations, numerical settings and final conclusions were developed, verified and approved by the author, who takes full responsibility for the content.